\documentclass[12pt,a4paper]{article}
\usepackage{axodraw}
\usepackage{epsfig}

\let\log\ln

\advance\textheight by \topskip
\if@twoside \oddsidemargin -17pt \evensidemargin 00pt
\else \oddsidemargin 00pt \evensidemargin 00pt
\fi
\expandafter\ifx\csname mathrm\endcsname\relax\def\mathrm#1{{\rm #1}}\fi

\renewcommand{\theequation}{\thesection.\arabic{equation}}
\newcounter{saveeqn}

\makeatletter 
\@addtoreset{equation}{section}
\makeatother

\def\beq{\begin{equation}}
\def\eeq{\end{equation}}
\def\beqar{\begin{eqnarray}}
\def\eeqar{\end{eqnarray}}
\def\barr#1{\begin{array}{#1}}
\def\earr{\end{array}}
\def\bfi{\begin{figure}}
\def\efi{\end{figure}}
\def\btab{\begin{table}}
\def\etab{\end{table}}
\def\bce{\begin{center}}
\def\ece{\end{center}}

\def\nl{\nonumber\\}
\def\nln{\nonumber\\*[-1ex]\phantom{\fbox{\rule{0em}{2ex}}}}

\def\al{\alpha}

\def\ga{\gamma}
\def\de{\delta}

\def\la{\lambda}
\def\si{\sigma}
\def\Ga{\Gamma}
\def\De{\Delta}

\def\refeq#1{\mbox{(\ref{#1})}}

\def\reffi#1{\mbox{Fig.~\ref{#1}}}

\def\refse#1{\mbox{Sect.~\ref{#1}}}

\def\citere#1{\mbox{Ref.~\cite{#1}}}
\def\citeres#1{\mbox{Refs.~\cite{#1}}}

\def\solid{\raise.9mm\hbox{\protect\rule{1.1cm}{.2mm}}}
\def\dash{\raise.9mm\hbox{\protect\rule{2mm}{.2mm}}\hspace*{1mm}}
\def\dot{\rlap{$\cdot$}\hspace*{2mm}}

\def\solid{\raise.9mm\hbox{\protect\rule{12mm}{.2mm}}}
\def\dash{\raise.9mm\hbox{\protect\rule{1.6mm}{.2mm}}\hspace*{1mm}}
\def\dot{\raise.9mm\hbox{\protect\rule{0.8mm}{.2mm}}\hspace*{0.8mm}}
\def\dashdot{\raise.9mm\hbox{\protect\rule{.3mm}{.2mm}}\hspace*{.8mm}\raise.9mm\hbox{\protect\rule{1.3mm}{.2mm}}\hspace*{.8mm}}

\newcommand{\GeV}{\unskip\,\mathrm{GeV}}

\newcommand{\TeV}{\unskip\,\mathrm{TeV}}
\newcommand{\fba}{\unskip\,\mathrm{fb}}

\def\mathswitchr#1{\relax\ifmmode{\mathrm{#1}}\else$\mathrm{#1}$\fi}

\newcommand{\PV}{\mathswitch V}
\newcommand{\PX}{\mathswitch X}
\newcommand{\PW}{\mathswitchr W}
\newcommand{\Pw}{\mathswitchr w}
\newcommand{\PZ}{\mathswitchr Z}

\newcommand{\Pe}{\mathswitchr e}
\newcommand{\Ri}{\mathswitchr i}
\newcommand{\Pne}{\mathswitch \nu_{\mathrm{e}}}

\newcommand{\Pnebar}{\mathswitch \bar\nu_{\mathrm{e}}}

\newcommand{\Pd}{\mathswitchr d}
\newcommand{\PD}{\mathswitchr D}

\newcommand{\Pu}{\mathswitchr u}
\newcommand{\PU}{\mathswitchr U}
\newcommand{\Ps}{\mathswitchr s}
\newcommand{\Pb}{\mathswitchr b}
\newcommand{\Pc}{\mathswitchr c}
\newcommand{\Pt}{\mathswitchr t}
\newcommand{\Pq}{\mathswitchr q}
\newcommand{\Pep}{\mathswitchr {e^+}}
\newcommand{\Pem}{\mathswitchr {e^-}}

\newcommand{\PWp}{\mathswitchr {W^+}}
\newcommand{\PWm}{\mathswitchr {W^-}}

\newcommand{\Pp}{\mathswitchr {p}}

\def\mathswitch#1{\relax\ifmmode#1\else$#1$\fi}

\newcommand{\MW}{\mathswitch {M_\PW}}

\newcommand{\MZ}{\mathswitch {M_\PZ}}

\newcommand{\Mt}{\mathswitch {m_\Pt}}
\newcommand{\GW}{\mathswitch {\Gamma_\PW}}
\newcommand{\GZ}{\Gamma_{\PZ}}
\newcommand{\PL}{\mathswitch {P_\PL}}

\newcommand{\scrs}{\scriptscriptstyle}
\newcommand{\sw}{\mathswitch {s_{\scrs\PW}}}

\def\ie{i.e.\ }
\def\eg{e.g.\ }

\newcommand{\Oa}{\mathswitch{{\cal{O}}(\alpha)}}
\newcommand{\Oas}{\mathswitch{{\cal{O}}(\alpha_s)}}

\newcommand{\rL}{{\mathrm{L}}}

\newcommand{\rd}{{\mathrm{d}}}

\newcommand{\ri}{{\mathrm{i}}}

\newcommand{\M}{{\cal{M}}}

\newcommand{\EW}{\mathrm{EW}}
\newcommand{\CM}{\mathrm{CM}}

\newcommand{\Born}{\mathrm{Born}}

\newcommand{\NLO}{\mathrm{NLO}}
\newcommand{\fact}{{\mathrm{fact}}}
\newcommand{\nonfact}{{\mathrm{nf}}}

\newcommand{\cut}{\mathrm{cut}}

\newcommand{\virt}{\mathrm{virt}}

\newcommand{\DPA}{{\mathrm{DPA}}}

\newcommand{\qqVVffff}{\bar q_1 q_2\to V_1 V_2 \to 4f}

\newcommand{\sparton}{\hat s}
\newcommand{\uparton}{\hat u}
\newcommand{\tparton}{\hat t}

\newcommand{\PT}{P_{\mathrm{T}}}
\newcommand{\ET}{E_{\mathrm{T}}}
\newcommand{\PTmiss}{P_{\mathrm{T}}^{\mathrm{miss}}}
\newcommand{\PTmax}{P_{\mathrm{T}}^{\mathrm{max}}}

\newcommand{\MT}{M_{\mathrm{T}}}

\newcommand{\Minv}{M_{\mathrm{inv}}}

\newcommand{\mr}[1]{\mathrm{#1}}

\makeatletter
\newcount\@tempcntc
\def\@citex[#1]#2{\if@filesw\immediate\write\@auxout{\string\citation{#2}}\fi
  \@tempcnta\z@\@tempcntb\m@ne\def\@citea{}\@cite{\@for\@citeb:=#2\do
    {\@ifundefined
       {b@\@citeb}{\@citeo\@tempcntb\m@ne\@citea
        \def\@citea{,\penalty\@m\ }{\bf ?}\@warning
       {Citation `\@citeb' on page \thepage \space undefined}}%
    {\setbox\z@\hbox{\global\@tempcntc0\csname
b@\@citeb\endcsname\relax}%
     \ifnum\@tempcntc=\z@ \@citeo\@tempcntb\m@ne
       \@citea\def\@citea{,\penalty\@m}
       \hbox{\csname b@\@citeb\endcsname}%
     \else
      \advance\@tempcntb\@ne
      \ifnum\@tempcntb=\@tempcntc
      \else\advance\@tempcntb\m@ne\@citeo
      \@tempcnta\@tempcntc\@tempcntb\@tempcntc\fi\fi}}\@citeo}{#1}}

\def\@citeo{\ifnum\@tempcnta>\@tempcntb\else\@citea
  \def\@citea{,\penalty\@m}%
  \ifnum\@tempcnta=\@tempcntb\the\@tempcnta\else
   {\advance\@tempcnta\@ne\ifnum\@tempcnta=\@tempcntb \else
\def\@citea{--}\fi
    \advance\@tempcnta\m@ne\the\@tempcnta\@citea\the\@tempcntb}\fi\fi}
\makeatother

\def\draftdate{\relax}
\def\mpar#1{\relax}
\def\mda{\relax}
\def\mua{\relax}
\def\mla{\relax}
\def\draft{
\def\thtystars{******************************}
\def\sixtystars{\thtystars\thtystars}
\typeout{}
\typeout{\sixtystars**}
\typeout{* Draft mode!
         For final version remove \protect\draft\space in source file *}
\typeout{\sixtystars**}
\typeout{}
\def\draftdate{\today}
\def\mua{\marginpar[\boldmath\hfil$\uparrow$]%
                   {\boldmath$\uparrow$\hfil}%
                    \typeout{marginpar: $\uparrow$}\ignorespaces}
\def\mda{\marginpar[\boldmath\hfil$\downarrow$]%
                   {\boldmath$\downarrow$\hfil}%
                    \typeout{marginpar: $\downarrow$}\ignorespaces}
\def\mla{\marginpar[\boldmath\hfil$\rightarrow$]%
                   {\boldmath$\leftarrow $\hfil}%
                    \typeout{marginpar: $\leftrightarrow$}\ignorespaces}
\def\Mua{\marginpar[\boldmath\hfil$\Uparrow$]%
                   {\boldmath$\Uparrow$\hfil}%
                    \typeout{marginpar: $\Uparrow$}\ignorespaces}
\def\Mda{\marginpar[\boldmath\hfil$\Downarrow$]%
                   {\boldmath$\Downarrow$\hfil}%
                    \typeout{marginpar: $\Downarrow$}\ignorespaces}
\def\Mla{\marginpar[\boldmath\hfil$\Rightarrow$]%
                   {\boldmath$\Leftarrow $\hfil}%
                    \typeout{marginpar: $\Leftrightarrow$}\ignorespaces}
\def\mpar##1{\marginpar{\hbadness10000%
                      \sloppy\hfuzz10pt\boldmath\bf##1}%
                      \typeout{marginpar: ##1}\ignorespaces}
\overfullrule 5pt
\oddsidemargin -15mm
\marginparwidth 29mm
}

\newcommand{\thismonth}{\ifcase\month\or January\or February\or March \or April
\or May \or June \or July \or August \or September \or \November \or 
\December\fi}

\makeatletter

\def\eqnarray{\stepcounter{equation}\let\@currentlabel=\theequation
\global\@eqnswtrue
\global\@eqcnt\z@\tabskip\@centering\let\\=\@eqncr
$$\halign to \displaywidth\bgroup\hskip\@centering
  $\displaystyle\tabskip\z@{##}$\@eqnsel&\global\@eqcnt\@ne
  \hskip 2\arraycolsep \hfil${##}$\hfil
  &\global\@eqcnt\tw@ \hskip 2\arraycolsep $\displaystyle\tabskip\z@{##}$\hfil
   \tabskip\@centering&\llap{##}\tabskip\z@\cr}
\def\appendix{\par
 \setcounter{section}{0} \setcounter{subsection}{0}
 \def\thesection{\Alph{section}}}

\makeatother

\begin{document}

\tolerance=100000
\thispagestyle{empty}
\setcounter{page}{0}

\thispagestyle{empty}
\def\thefootnote{\fnsymbol{footnote}}
\setcounter{footnote}{1}
\null
\draftdate\hfill  PSI-PR-05-10
\\
\strut\hfill ZU-TH 19/05 \\
\strut\hfill DFTT 36/05\\
\strut\hfill hep-ph/0511088
\vskip 0cm
\vfill
\begin{center}
  {\Large \bf Electroweak corrections and anomalous triple gauge-boson 
   couplings in $W^+W^-$ and $W^\pm Z$ production at the LHC
\par} \vskip 2.5em
{\large
{\sc E. Accomando$^1$ and A. Kaiser$^{2,3}$}}%
\\[.5cm]
$^1$ {\it Dipartimento di Fisica Teorica, Universit\`a di Torino,\\
Via P. Giuria 1, 10125 Torino, Italy}
\\[0.3cm]
$^2$ {\it Paul Scherrer Institut\\
CH-5232 Villigen PSI, Switzerland}
\\[0.3cm]
$^3$ {\it Institute of Theoretical Physics\\ University of Z\"urich, CH-8057 
Z\"urich, Switzerland}
\par
\end{center}\par
\vskip 2.0cm \vfill {\bf Abstract:} \par 
We have analysed the production of $\PW\PZ$ and $\PW\PW$ vector-boson pairs at
the LHC. These processes give rise to four-fermion final states, and are
particularly sensitive to possible non-standard trilinear gauge-boson 
couplings. We have studied the interplay between the influence of these 
anomalous couplings and the effect of the complete logarithmic electroweak 
$\Oa$ corrections. Radiative corrections to the Standard Model processes 
in double-pole approximation and non-standard terms due to trilinear 
couplings are implemented into a Monte Carlo program for 
$\Pp\Pp\to 4f (+\gamma )$ with final states involving four or two charged 
leptons. We numerically investigate purely leptonic final 
states and find that electroweak corrections can fake new-physics signals, 
modifying the observables by the same amount and shape, in kinematical
regions of statistical significance.   
\par
\vskip 1cm
\noindent
November 2005
\par
\null
\setcounter{page}{0}
\clearpage
\def\thefootnote{\arabic{footnote}}
\setcounter{footnote}{0}

\def\mla{\marginpar[\boldmath\hfil$\rightarrow$]%
                   {\boldmath$\leftarrow $\hfil}%
                    \typeout{marginpar: $\leftrightarrow$}\ignorespaces}
\def\mua{\marginpar[\boldmath\hfil$\uparrow$]%
                   {\boldmath$\uparrow$\hfil}%
                    \typeout{marginpar: $\uparrow$}\ignorespaces}
\def\mda{\marginpar[\boldmath\hfil$\downarrow$]%
                   {\boldmath$\downarrow$\hfil}%
                    \typeout{marginpar: $\downarrow$}\ignorespaces}

\section{ Introduction }
\label{sec:intro}

In the last few years, LEP2 and Tevatron have provided accurate tests of the 
non-abelian structure of the Standard Model (SM), probing the existence of 
self-interactions among electroweak gauge bosons. The experimental 
collaborations have performed several measurements of charged and neutral 
triple gauge-boson couplings (TGCs), mainly analysing the production of 
gauge-boson 
pairs whose cross sections depend very sensitively on the non-abelian sector 
of the underlying theory. Still, up to now the self-couplings have not been 
determined with the same precision as other boson properties, such as their 
masses and couplings to fermions. Despite the copious production of 
$\PWp\PWm$ pairs at LEP2, the experimental bounds on possible
anomalous couplings, which parametrize deviations from SM predictions
due to new physics occurring at energy scales of order of tens of TeV,
are not very stringent. The weakness of the LEP2 measurement is the rather 
modest energy scale at which $\PW$-pair-production events have been generated.
Anomalous gauge-boson couplings are in fact expected to increasingly enhance 
the gauge-boson pair-production cross section at large di-boson invariant 
masses $M_{VV^\prime}$ ($V,V^\prime =\PW,\PZ,\gamma$), as they spoil the 
unitarity cancellations for longitudinal gauge bosons. Hence, at future 
colliders it will be useful to analyse the di-boson production at the highest 
possible center-of-mass (CM) energies.

In the near future, the Large Hadron Collider (LHC) will be the main source 
of vector-boson pairs produced with large invariant mass $M_{VV^\prime}$.
The machine will collect hundred thousands of events, the exact statistics
depending on the particular process and luminosity \cite{Haywood:1999qg}. 
The prospects for a detailed investigation of trilinear couplings will
sensibly improve when the envisaged integrated luminosity of $100\fba^{-1}$
will be reached. Owing to the expected increase in statistics, the 
measurement of anomalous TGCs requires theoretical 
predictions from Monte Carlo generators of order of a few per cent accuracy 
to allow for a decent data analysis. At lowest order, this means taking into
account spin correlation and finite-width effects, as well as the 
contribution of the irreducible background coming from all Feynman diagrams 
which are not mediated by di-boson production but give rise to the same final 
state. Whenever dominant, these diagrams could spoil the sensitivity to 
possible new physics, as they do not contain triple gauge-boson couplings. 
The way to achieve this level of precision is to compute the complete process 
$\Pp\Pp\to 4f$, going beyond the {\it{production$\times$decay}} approach. 
This represents the most basic step towards the desired accuracy. Moreover, a 
full understanding and control of higher order QCD and electroweak ($\EW$) 
corrections is necessary to match the experimental error.

In the past years, hadronic di-boson production has been studied 
extensively by many authors, with particular attention to the $\Oas$ QCD 
corrections (for a review on the subject see \citere{Haywood:1999qg}). 
Several NLO Monte Carlo programs have been constructed and cross checked so 
that complete $\Oas$ corrections are now available 
\cite{Dixon:1999di,DeFlorian:2000sg,Campbell:1999ah}. Inclusive NLO QCD 
corrections turn out to be very large at LHC energies. They can increase the 
overall lowest-order cross section by a factor of two, if no cuts are applied.
Their effect is even more pronounced if one considers kinematical 
distributions particularly sensitive to anomalous couplings. 
As an example, QCD corrections can increase the tails of vector-boson 
transverse momentum and di-boson invariant mass distributions by one order of 
magnitude \cite{Ohnemus:1991gb,Frixione:1992pj}, thus spoiling the sensitivity
to possible deviations from the SM.  By including a jet veto, their effects 
are drastically reduced to the order of tens of per cent 
\cite{Dixon:1999di,Baur:1995aj}, restoring the sensitivity to anomalous 
$\PW\PW\PV$ couplings. 

In view of the envisaged precision of a few per cent at the LHC, also a 
discussion of $\EW$ corrections is in order (see for example 
\citere{Hollik:2004} and references therein). 
Various analyses of the effect of one-loop logarithmic $\EW$ corrections on 
$\PW\ga$, $\PZ\ga$, $\PW\PZ$ and $\PW\PW$ production processes at the LHC 
\cite{Accomando:2001fn,Kaiser:2004,Meier:2005} have pointed out that $\Oa$ 
corrections are comparable or bigger than the statistical error, 
when exploring large di-boson invariant masses and large rapidity of the 
produced gauge-bosons. This is precisely the kinematical region where effects 
due to anomalous couplings are expected to be maximally enhanced.
Hence, for a meaningful analysis of possible new-physics effects in high 
energy domains of suitable distributions, including only universal
radiative corrections such as the running of the electromagnetic
coupling and corrections to the $\rho$ parameter is not enough.
The remaining $\EW$ corrections, enhanced by double and single logarithms
of the ratio of the CM-energy to the $\EW$ scale, may be indeed relevant.  
The growth of $\Oa$ $\EW$ corrections with increasing energy is well 
known since long time. Analyses of the general high-energy behaviour of $\EW$
corrections have been extensively performed (see for instance
\citeres{Beenakker:1993tt,ewee}). 
From the computational point of view, a process-independent recipe greatly 
simplifies the calculation of leading-logarithmic $\EW$ corrections. Such a  
method is described in \citeres{Denner:2001jv,Denner:2001gw}. 
There, it has been shown that the leading-logarithmic one-loop corrections to 
arbitrary $\EW$ processes factorize into the tree-level amplitudes times 
universal correction factors.

Using the method of \citeres{Denner:2001jv,Denner:2001gw}, we have 
investigated in \citere{Kaiser:2004} the effect of leading-logarithmic $\Oa$ 
$\EW$ corrections to the hadronic production of $\PW^\pm \PZ$ and 
$\PW^\pm \PW^\mp$ 
pairs in the large-invariant-mass region of the hard process at the LHC.  
In this paper, we compare their shape and size with the influence of 
anomalous TGCs on the lowest-order SM predictions. In this study, QCD 
corrections are not included.
The simplest experimental analyses of gauge-boson pair production will
rely on purely leptonic final states. Semi-leptonic channels, where
one of the vector bosons decays hadronically, have been analysed at
the Tevatron \cite{semitev} showing that these events suffer from the
background due to the production of one vector boson plus jets via
gluon exchange.  For this reason, we study only di-boson production where 
both gauge bosons decay leptonically into $\Pe$ or $\mu$.

The paper is organized as follows: the relevant triple gauge-boson couplings 
and the parametrization used to calculate their contribution to 
$pp\rightarrow 4f(+\gamma)$ processes 
are given in \refse{sec:tgc}. The strategy of our calculation, which improves 
the tree-level predictions by including one-loop electroweak corrections, is
described in \refse{sec:strategy-calculation}. The general setup of
our numerical analysis and the discussion of processes mediated by $\PW\PZ$ 
and $\PW\PW$ production are given in \refse{sec:studies}. Our findings
are summarized in \refse{sec:concl}.  

\section{Triple gauge-boson couplings}
\label{sec:tgc}

New physics occurring at energy scales much larger than those probed
directly at forthcoming experiments could modify the structure of the
vector-boson self-interactions. These modifications are parametrized in terms
of anomalous couplings in the Yang-Mills vertices.
The hadronic production of $\PW\PW$ and $\PW\PZ$ pairs is sensitive to possible
anomalous triple gauge-boson couplings in the charged sector, \ie  
to anomalous $\PW^+\PW^-\PZ$ and $\PW^+\PW^-\gamma$ couplings\footnote{We do 
not discuss here purely neutral gauge-boson couplings, involving only $\PZ$ 
and $\gamma$.}.
The two most general vertices, which preserve Lorentz invariance and separate 
C- and P-conservation, are described by the effective Lagrangian suggested in
\citere{effective_lagrangian}:
\beq
\label{eq:lagrangian}
\mathcal{L}_{WWV}=g_{\PW\PW\PV}\left [g_1^\PV(\PW^\dagger_{\mu\nu}\PW^\mu\PV^\nu -
\PW^\dagger_\mu\PV_\nu\PW^{\mu\nu})+\kappa^\PV\PW^\dagger_\mu\PW_\nu\PV^{\mu\nu}
+{\la^\PV\over{\MW^2}}\PW^\dagger_{\rho\mu}\PW^\mu_\nu\PV^{\nu\rho}\right ]
\eeq
where $\PV^\mu$ represents the $\PZ$ and $\gamma$ fields, 
$\PX^{\mu\nu}=\partial_\mu\PX_\nu -\partial_\nu\PX_\mu$ 
(for $\PX=\PW,\PZ,\gamma$), 
$g_{\PW\PW\gamma}=-e$ and $g_{\PW\PW\PZ}= e\cot\theta_\Pw$, 
with $\theta_\Pw$ the weak mixing angle and $e$ the electric charge.
For simplicity, C- or P-violating $\PW\PW\PV$ couplings are not considered 
in this paper.
The six free parameters in eq.\refeq{eq:lagrangian} can be written in terms 
of their deviation, $\De$, from the corresponding SM values:
\beq
g_1^\PV=1+\De g_1^\PV,~~~~~~~\kappa^\PV=1+\De \kappa^\PV,~~~~~~~\la^\PV=\De\la^\PV.
\eeq
Instead of using rather general parametrizations of non-standard couplings, 
we adopt a convention commonly used in the LEP2 data analysis 
\cite{lep2tgcconvention} to reduce the number of free parameters. We assume 
in the following that $\De g_1^\gamma =0$. The remaining couplings are 
further constrained by the relations:
\begin{equation}
 \lambda_\PZ = \lambda_\gamma, \qquad \De\kappa_\PZ = \De g_1^\PZ - \tan^2 \theta_\Pw \De\kappa_\gamma .
\end{equation}
In this approach, we are thus left with only three independent parameters, 
\ie $g_1^\PZ$, $\kappa_\gamma$, and $\lambda_\gamma$.
LEP2 and Tevatron have constrained the value of the $\PW\PW\PV$ coupling
constants at the few-per-cent level. The experimental average gives the 
following 95$\%$ confidence intervals \cite{pdg}:
\beq
-0.054\le\De g_1^\PZ\le 0.028,~~~~~~-0.117\le\De\kappa_\gamma\le 0.061,~~~~~~
-0.07\le\De \lambda_\gamma\le 0.012,
\eeq
where each parameter has been determined from a single-parameter fit, that is
performed by assuming SM values for all other couplings. 
Taking constant values for the anomalous couplings in the effective 
Lagrangian \refeq{eq:lagrangian} would violate unitarity. In order to 
preserve that, any deviation from the SM expectations must be inserted into
the vertices via a form factor which vanishes at asymptotically high energies 
\cite{formfactor}:
\beq\label{eq:formfactor}
\De^\prime Y= {\De Y\over{(1+\hat{s}/\Lambda^2_{FF})^n}}~,~~~~~~~~~Y=g_1^\PZ, 
\kappa_\gamma ,\lambda_\gamma , 
\eeq
with $\De Y$ the value at low energy, $\sqrt{\hat{s}}$ the partonic $\CM$ 
energy, and $\Lambda_{FF}$ the energy scale at which new physics could 
possibly appear. 

At the Born level, it is straightforward to include anomalous couplings in 
the matrix elements. On the contrary, at one-loop, non-standard model 
contributions do not guarantee the renormalizability of the electroweak 
theory. Consequently, we consider their effect only on the lowest-order cross
section. 

\section{Strategy of the calculation}
\label{sec:strategy-calculation}

We consider the production of massive gauge-boson pairs in proton-proton
collisions. In the parton model the corresponding cross sections are 
described by the following convolution
\beqar\label{eq:convolution}
\rd\si^{h_1 h_2}(P_1,P_2,p_f) = \sum_{i,j}\int_0^1\rd x_1 \int_0^1\rd x_2\,
\Phi_{i,h_1}(x_1,Q^2)\Phi_{j,h_2}(x_2,Q^2) \,
\rd\hat\si^{ij}(x_1P_1,x_2P_2,p_f),\nln
\eeqar
\begin{figure}
\begin{center}
\begin{picture}(400,200)(0,0)

\SetOffset(0,0)

\Line(50,160)(100,160)
\Line(50,165)(100,165)
\Line(50,155)(100,155)
\ArrowLine(150,100)(100,160)
\LongArrow(100,160)(150,190)
\LongArrow(100,162)(145,195)
\LongArrow(100,158)(155,185)

\Line(50,40)(100,40)
\Line(50,45)(100,45)
\Line(50,35)(100,35)
\ArrowLine(100,40)(150,100)
\LongArrow(100,40)(150,10)
\LongArrow(100,42)(155,15)
\LongArrow(100,38)(145,5)

\GCirc(100,40){12}{0.5}
\GCirc(100,160){12}{0.5}

\Photon(150,100)(250,50){3}{10}
\Photon(150,100)(250,150){-3}{10}
\GCirc(150,100){15}{0.5}

\DashLine(180,10)(180,190){5}
\DashLine(230,10)(230,190){5}
\GCirc(205,125){10}{1.0}
\GCirc(205,75){10}{1.0}

\ArrowLine(300,180)(250,150)
\ArrowLine(250,150)(300,120)
\ArrowLine(300,80)(250,50)
\ArrowLine(250,50)(300,20)

\GCirc(250,150){12}{0.5}
\GCirc(250,50){12}{0.5}

\Text(50,180)[]{Proton}
\Text(50,20)[]{Proton}

\Text(135,145)[]{$\bar q_1$}
\Text(135,55)[]{$q_2$}

\Text(310,180)[]{$\bar f_4$}
\Text(310,120)[]{$f_3$}
\Text(310,80)[]{$\bar f_6$}
\Text(310,20)[]{$f_5$}

\Text(200,145)[]{$V_1$}
\Text(200,55)[]{$V_2$}

\end{picture}
\end{center}

\caption{Structure of the  process $\Pp\Pp\to V_1 V_2+X \to 4f+X$ in $\DPA$}
\label{Full_Amplitude_DPA}
\end{figure}
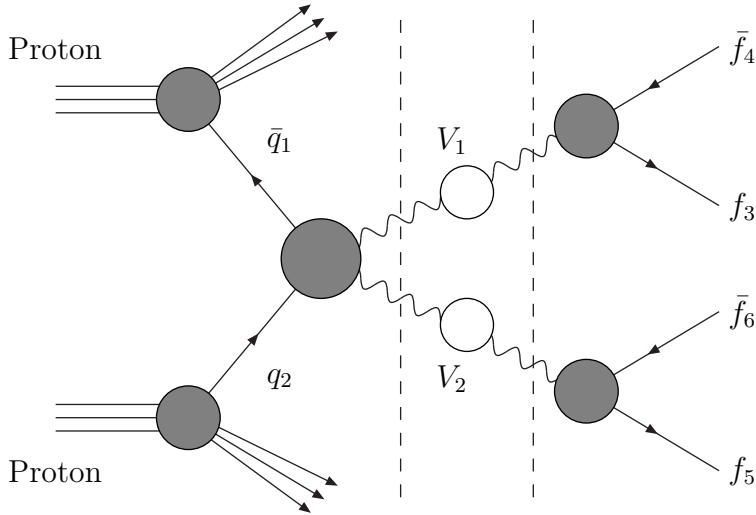
\par\noindent
where $p_f$ summarizes the final-state momenta, $\Phi_{i,h_1}$ and 
$\Phi_{j,h_2}$ are the distribution functions of the partons $i$ and $j$ in 
the incoming protons $h_1$ and $h_2$ with momenta $P_1$ and $P_2$, 
respectively, $Q$ is the factorization scale, and $\hat\si^{ij}$ represent 
the cross sections for the partonic processes averaged over colours and spins 
of the partons. At lowest-order, these cross sections are calculated using 
the matrix elements for the complete process
\beq\label{eq:process}
\bar q_1(p_1,\sigma_1) + q_2(p_2,\sigma_2)\rightarrow f_3(p_3,\sigma_3)
+ f_4(p_4,\sigma_4) + f_5(p_5,\sigma_5) + f_6(p_6,\sigma_6) 
\eeq
where the arguments label momenta $p_i$ and helicities $\sigma_i$ of the
external fermions. This means that we include the full set of Feynman 
diagrams, in this way accounting for the resonant di-boson production as well 
as the irreducible background coming from non-doubly resonant contributions.
Complete four-fermion phase spaces and exact kinematics are employed in our
calculation. For the evaluation of the electroweak corrections we follow the 
approach developed and described in \citeres{Kaiser:2004,Denner:2000bj}. 
Explicit formulas for the processes \refeq{eq:process} discussed in this 
paper are given in \citere{Kaiser:2004}. In the following we simply summarize 
the kernel of the adopted approximations and discuss their applicability 
domains.

The virtual corrections, coming from loop diagrams, are computed in 
double-pole approximation ($\DPA$), that is taking into account only those 
terms which are enhanced by two resonant gauge-boson propagators, 
$\bar q_1 q_2\rightarrow V_1V_2\rightarrow 4f$. 
In $\DPA$, the generic process we want to analyse has the structure depicted 
in Fig. 1. The matrix element factorizes into the production of two 
on-shell bosons, $\M_{\Born}^{\bar q_1 q_2\to V_{1,\la_1}V_{2,\la_2}}$, their
propagators, and their decay into fermion pairs, 
$\M_\Born^{V_{k,\la_k}\to f_i\bar f_j}$, 
\beqar\label{eq:BornVV} \M_{\Born,\DPA}^{\qqVVffff} &=&
P_{V_1}(k_1^2)\;P_{V_2}(k_2^2)
  \sum_{\la_1,\la_2}\M_{\Born}^{\bar q_1q_2\to
  V_{1,\la_1}V_{2,\la_2}} \M_\Born^{V_{1,\la_1}\to
  f_3\bar f_4}\M_\Born^{V_{2,\la_2}\to f_5\bar f_6}.
\eeqar 
The sum runs over the physical helicities $\la_1,\la_2=0,\pm1$ of the
on-shell projected gauge bosons $V_1$ and $V_2$ with momenta $k_1$ and
$k_2$, respectively.  The propagators of the massive gauge bosons 
\begin{equation}\label{eq:PV}
 P_V(p) = \frac{1}{p^2 - M_V^2+\theta(p^2)\ri M_V\Ga_V}, 
\quad V=W,Z
\end{equation}
involve besides the masses of the gauge bosons also their widths, which we
consider as constant and finite for time-like momenta. In this approximation, 
the $\Oa$ virtual corrections are of two types: factorizable and 
non-factorizable ones. The former are those that can be associated either
to the production or to the decay subprocess. Their matrix elements for the 
processes $\qqVVffff$ can be written as
\beqar\label{eq:factcorrVV}
\de\M_{\virt,\DPA,\fact}^{\qqVVffff} &=&
P_{V_1}(k_1^2)\;P_{V_2}(k_2^2)
\sum_{\la_1,\la_2}\biggl\{\de\M_{\virt}^{\bar q_1q_2\to V_{1,\la_1}V_{2,\la_2}}
\M_{\Born}^{V_{1,\la_1}\to f_3\bar f_4}\M_{\Born}^{V_{2,\la_2}\to f_5\bar f_6}\nl
&&{}+\M_{\Born}^{\bar q_1q_2\to V_{1,\la_1}V_{2,\la_2}}
\de\M_{\virt}^{V_{1,\la_1}\to f_3\bar f_4}\M_{\Born}^{V_{2,\la_2}\to f_5\bar f_6}\nl
&&{}+\M_{\Born}^{\bar q_1q_2\to V_{1,\la_1}V_{2,\la_2}}
\M_{\Born}^{V_{1,\la_1}\to f_3\bar f_4}\de\M_{\virt}^{V_{2,\la_2}\to f_5\bar f_6}\biggr\},
\eeqar
where $\delta\M_{\virt}^{\bar q_1q_2\to V_{1,\la_1}V_{2,\la_2}}$,
$\delta\M_\virt^{V_{1,\la_1}\to  f_3\bar f_4}$, and
$\delta\M_\virt^{V_{2,\la_2}\to  f_5\bar f_6}$ denote the virtual corrections 
to the on-shell matrix elements for the gauge-boson production and decay 
processes. The latter ones connect instead production and decay subprocesses 
or two decay subprocesses, and yield a simple correction factor 
$\de^{\virt}_{\nonfact,\DPA}$ to the lowest-order cross section. 

We calculate factorizable and non-factorizable $\Oa$ virtual corrections in 
logarithmic high-energy approximation, taking into account only contributions 
involving single and double enhanced logarithms at high energies, \ie $\Oa$ 
contributions proportional to $\al\log^2(|\hat s|/\MW^2)$ or 
$\al\log(|\hat s|/\MW^2)$, where $\sqrt{\hat s}$ is the CM-energy of the 
partonic subprocess. The logarithmic approximation yields the dominant 
corrections as long as CM-energies and scattering angles are large. 
Pure angular-dependent logarithms of the form $\al\log^2(|\hat s|/\hat r)$ or 
$\al\log(|\hat s|/\hat r)$, with $\hat r$ equal to the Mandelstam variables
$\hat t$ and $\hat u$ of the partonic production subprocess, are in fact not 
included. The validity of the results relies therefore on the assumption that 
all invariants are large compared with $\MW^2$ and approximately of the same 
size
\beq\label{HEA}
{\sparton}\sim |{\tparton}|\sim |{\uparton}|\gg \MW^2.
\eeq
This implies that the produced gauge bosons should be energetic and emitted 
at sufficiently wide angles with respect to the beam. This is precisely the 
kinematical region where effects due to possible anomalous couplings should 
be most enhanced. In this region, the accuracy of the logarithmic high-energy
approximation is expected to be of order of a few per cent. Numerical 
estimates of the omitted terms, based on the comparison 
between complete $\Oa$ corrections and their high-energy limit for different
processes \cite{Meier:2005,Beenakker:1993tt}, confirm this level of 
precision. We can 
thus reasonably adopt this approximation at the LHC, where the experimental
error in the high-energy regime is at the few-per-cent level.

The afore-mentioned $\Oa$ contributions originate from above the $\EW$ scale,
and affect only the production subprocess. In addition, one has to consider 
purely electromagnetic logarithmic corrections of the form 
$\log(\MW^2/m_f^2)$ or $\log(\MW^2/\la^2)$, where $\la$ is the photon mass 
regulator and $m_f$ the fermion mass, which originate from below the $\EW$ 
scale. These large logarithms from diagrams with photon exchange affect also 
the decay subprocesses, giving rise to a correction factor proportional to 
the lowest-order matrix element \cite{Kaiser:2004}.
\par\noindent
Soft and collinear singularities, must be cancelled against their 
counterparts in the real corrections. Conversely to the virtual corrections,
these latter ones are calculated using the matrix elements for the complete 
processes 
\beq
\bar q_1(p_1,\sigma_1) + q_2(p_2,\sigma_2)\rightarrow f_3(p_3,\sigma_3)
+ f_4(p_4,\sigma_4) + f_5(p_5,\sigma_5) + f_6(p_6,\sigma_6) + 
\gamma (k,\lambda_\gamma ) 
\eeq
with emission of an additional photon of momentum $k$ and helicity 
$\lambda_\gamma = \pm 1$. The well-known phase-space slicing method 
(see \eg \citere{Berends:1981uq}) is employed for isolating soft and collinear 
divergencies. The details of the implementation are given in 
\citere{Kaiser:2004}. 

\section{Numerical studies}
\label{sec:studies}

In this section, we illustrate the impact of the one-loop electroweak 
radiative corrections on the observability of anomalous triple gauge-boson 
couplings in $\PW\PZ$ and $\PW\PW$ production at the LHC. 
We consider two classes of processes:
\renewcommand{\labelenumi}{(\roman{enumi})}
\begin{enumerate}
\item $\Pp\Pp\to l\nu_ll^\prime\bar{l^\prime}($+$\gamma )$,
\qquad 
\item $\Pp\Pp\to l\bar\nu_l\nu_{l^\prime}\bar{l^\prime}(+\gamma )$, 
\end{enumerate}
where $l,l^\prime=\Pe$ or $\mu$. In our notation, $l\nu_l$ indicates
both $l^-\bar\nu_l$ and $l^+\nu_l$.  The first class is characterized
by three isolated charged leptons plus missing energy in the final
state.  This channel includes $\PW\PZ$ production as intermediate
state.  The second class is instead related to $\PW^\pm\PW^\mp$ production. 
When there is a unique flavor in the final state, $l=l'$, the latter
process receives also a $\PZ\PZ$ contribution. 
In the parton model the corresponding cross sections are described by the
convolution in eq.\refeq{eq:convolution}.
Since the two incoming hadrons are protons and we sum over final
states which are related one another by charge conjugation, we find
\beqar\label{eq:convolWZ}
\rd\si^{\Pp\Pp}(P_1,P_2,p_f) = 
\int_0^1\rd x_1 \rd x_2 &&\sum_{U=\Pu,\Pc}\sum_{D=\Pd,\Ps}
\Bigl[\Phi_{\bar\PD,\Pp}(x_1,Q^2)\Phi_{\PU,\Pp}(x_2,Q^2)\,\rd\hat\si^{\bar\PD\PU}
(x_1P_1,x_2P_2,p_f)
\nl&&{}
+\Phi_{\bar\PU,\Pp}(x_1,Q^2)\Phi_{\PD,\Pp}(x_2,Q^2)\,\rd\hat\si^{\bar\PU\PD}
(x_1P_1,x_2P_2,p_f)
\nl&&{}
+\Phi_{\bar\PD,\Pp}(x_2,Q^2)\Phi_{\PU,\Pp}(x_1,Q^2)\,\rd\hat\si^{\bar\PD\PU}
(x_2P_2,x_1P_1,p_f)
\nl&&{}
+\Phi_{\bar\PU,\Pp}(x_2,Q^2)\Phi_{\PD,\Pp}(x_1,Q^2)\,\rd\hat\si^{\bar\PU\PD}
(x_2P_2,x_1P_1,p_f)
\Bigr]
\eeqar
for $\PW\PZ$ production and 
\beqar\label{eq:convolVV}
\rd\si^{\Pp\Pp}(P_1,P_2,p_f) = 
\int_0^1\rd x_1 \rd x_2 &&\sum_{q=\Pu,\Pd,\Pc,\Ps}
\Bigl[\Phi_{\bar\Pq,\Pp}(x_1,Q^2)\Phi_{\Pq,\Pp}(x_2,Q^2)\,\rd\hat\si^{\bar\Pq\Pq}
(x_1P_1,x_2P_2,p_f)
\nl&&{}
+\Phi_{\bar\Pq,\Pp}(x_2,Q^2)\Phi_{\Pq,\Pp}(x_1,Q^2)\,\rd\hat\si^{\bar\Pq\Pq}
(x_2P_2,x_1P_1,p_f)
\Bigr]
\eeqar
for $\PW\PW$ (and $\PZ\PZ$) production in leading order of QCD. In computing
partonic cross-sections, for the free parameters we use the input values 
\cite{Hagiwara:pw,mtop}:
\beq\label{eq:SMpar}
\begin{array}[b]{lcllcllcl}
G_\mu &= & 1.16637 \times 10^{-5} \GeV^{-2}, \qquad &
\MW & = & 80.425\GeV, \qquad &
\MZ & = & 91.1876\GeV, \qquad  \\
\Mt & = & 178.0 \GeV.  \quad \\
\end{array}
\eeq
The weak mixing angle is fixed by $\sw^2=1-\MW^2/\MZ^2$.  Moreover, we
adopted the so-called $G_{\mu}$-scheme, which effectively includes
higher-order contributions associated with the running of the
electromagnetic coupling and the leading universal two-loop
$\Mt$-dependent corrections. To this end we parametrize the
lowest-order matrix element in terms of the effective coupling
$\alpha_{G_{\mu}}=\sqrt{2}G_{\mu}\MW^2\sw^2/\pi 
= 7.543596\ldots \times 10^{-3}$  and omit the explicit contributions 
proportional to $\De\al(\MW^2)$ and $\De\al(\MZ^2)$ in the electroweak 
virtual corrections due to parameter renormalization.
Additional inputs are the quark-mixing matrix elements whose
values have been taken to be $V_{\Pu\Pd}=0.974$ \cite{Hocker:2001xe},
$V_{\Pc\Ps}=V_{\Pu\Pd}$, $V_{\Pu\Ps}= -V_{\Pc\Pd}= \sqrt{1-
  |V_{\Pu\Pd}|^2} = 0.226548\ldots$, $V_{\Pt\Pb}=1$, and zero for all other 
matrix elements. We have moreover used the fixed-width scheme with 
$\GZ = 2.505044 \GeV$ and $\GW = 2.099360 \GeV$. As to parton distributions, 
we have chosen CTEQ6M \cite{cteq} at the following factorization scales:
\begin{equation}\label{eq:scaleWZ}
Q^2={1\over 2}\left (\MW^2+\MZ^2+\PT^2(l\nu_l)+
\PT^2(l^\prime\bar{l^\prime})\right )
\end{equation}
and 
\begin{equation}\label{eq:scaleWW}
Q^2={1\over 2}\left (2\MW^2+\PT^2(l)+\PT^2(l^\prime )+
\PT^2(\nu\nu^\prime )\right ) .
\end{equation}
for $\PW\PZ$ and $\PW\PW$ production processes, respectively, where $\PT$
is the transverse momentum.
For final states that allow for two different sets of reconstructed
gauge bosons, we choose the average of the corresponding scales from 
\refeq{eq:scaleWZ}--\refeq{eq:scaleWW} if both reconstructed sets pass
the cuts.
This scale choice appears to be appropriate for the calculation of
differential cross sections, in particular for vector-boson
transverse-momentum distributions. It generalizes the scale of
\citeres{Frixione:1992pj,Dixon:1999di} to final states with identical
particles.
\par\noindent
For the experimental identification of the final states to be analysed, 
we have implemented a general set of cuts appropriate for LHC, and defined 
as follows:
\begin{itemize}
\item {charged lepton transverse momentum $\PT(l)>20\GeV$},
  
\item {missing transverse momentum $\PTmiss> 20\GeV$ 
for final states with one neutrino and  $\PTmiss> 25\GeV$
for final states with two neutrinos},
  
\item {charged lepton pseudo-rapidity $|\eta_l |< 3$}, where
  $\eta_l=-\log\left (\tan(\theta_l/2)\right )$, and $\theta_l$ is the
  polar angle of particle $l$ with respect to the beam.
\end{itemize}

\begin{table}\centering
\begin{tabular}{|c|c|c|c|c|c|}
\hline 
Scenario  & $\lambda_\gamma$ & $\lambda_\PZ$ & $\Delta g_1^\PZ$ & 
$\Delta \kappa_\gamma$ & $\Delta \kappa_\PZ$ \\ 
\hline
 Born & $0$ & $0$ & $0$ & $0$ & $0$ \\ 
 $2a/2b$ & $0$ & $0$ & $\pm 0.02$ & $0$ & $\pm 0.02$ \\ 
 $3a/3b$ & $0$ & $0$ & $0$ & $\pm 0.04$ & $\mp 0.01142 $ \\ 
 $4a/4b$ & $\pm 0.02$ & $\pm 0.02$ & $0$ & $0$ & $0$ \\ 
\hline
\end{tabular}
\caption {Different scenarios for the single-parameter analysis of
the anomalous triple gauge-boson couplings. Letters $a$ and $b$ correspond to
positive and negative values, respectively.}
\label{ta:tgc_scenarios}
\end{table}
\noindent
These cuts approximately simulate the detector acceptance. At Born level,
they can be directly implemented on the final state particles. A complication
arises at one-loop level. When calculating real-photonic corrections, the 
emission of an additional real photon must be taken into account. The
afore-mentioned acceptance cuts assume a perfect separation of this extra 
photon from the charged leptons, which is not very realistic. In order to give
a description of the final state closer to the experimental situation, we 
consider the following photon recombination procedure:
\begin{itemize}
\item {Photons with a rapidity $|\eta_{\gamma} |> 3$ are treated as
invisible}.

\item {If the photon is central enough ($|\eta_{\gamma} |< 3$) and 
the rapidity--azimuthal-angle separation between charged 
lepton and photon $\Delta R_{l\gamma}=
\sqrt{(\eta_l-\eta_\gamma )^2+(\phi_l -\phi_\ga)^2}<0.1$, then the photon 
and lepton momentum four-vectors are combined into an effective lepton
momentum}.

\item {If the photon is central enough ($|\eta_{\gamma} |< 3$), 
the rapidity--azimuthal-angle separation $\Delta R_{l\gamma}>0.1$,
and the photon energy $E_\gamma < 2\GeV$, then the momenta of the photon 
and of the nearest charged lepton are recombined}.

\item {In all other cases we assume that the photon can be distinguished in
the detector and therefore does not contribute to the processes in 
consideration. This last requirement amounts to a photon veto, as we discard
all events with a visible photon.}

\end{itemize}
Let us notice that this recombination procedure differs from the one adopted 
in \citere{Kaiser:2004}. The results presented in the following sections 
cannot be therefore directly compared with those of \citere{Kaiser:2004}.
After photon recombination, the effective lepton momentum must pass the 
acceptance cuts for the different processes, and we use effective lepton 
momenta to define the above-mentioned factorization scales. For the processes 
considered, we have also implemented further cuts which are described in due 
time. 

In the following sections, we present 
results for the LHC at $\CM$ energy $\sqrt s=14\TeV$ and an integrated 
luminosity $L=100\fba^{-1}$. We assume a dipole form factor ($n=2$) with 
scale $\Lambda_{FF}=1\TeV$ in eq.\refeq{eq:formfactor}. In order to study the 
effect of anomalous triple gauge-boson couplings, we perform a 
single-parameter analysis.
We thus vary one of the independent parameters $\lambda_\gamma$, 
$\Delta g_1^\PZ$, $\Delta \kappa_\gamma$ at a time, keeping the remaining 
ones at their SM zero value. The considered scenarios are summarized in 
Table 1, for some
representative values. The chosen numbers are meant to be a pure sample 
set. The purpose of this paper is not a realistic and exhaustive analysis of
the observability of new-physics effects. The aim is to give evidence on the 
interplay between non-standard terms and $\EW$ corrections in a realistic
context, \ie taking into account the present anomalous TGCs exclusion limits 
and the planned LHC potential.     
Nonetheless, our Monte Carlo could serve as a tool to estimate the full 
sensitivity of LHC to non-standard couplings via differential cross section 
studies and event selections.

\subsection{$\PW\PZ$ production}

In this section, we study the leptonic processes 
$\Pp\Pp\to l\nu_l l^\prime\bar{l^\prime}$ with $l,l^\prime = \Pe$ or $\mu$. 
These final states are relatively background free, and can be mediated by 
$\PW\PZ$ production. Hence, they provide a good testing ground for the 
trilinear $\PW\PW\PZ$ coupling, once the $\PZ$- and $\PW$ bosons are properly 
reconstructed. We simulate the $\PZ$-boson selection by requiring
at least one pair of opposite-sign leptons with invariant mass satisfying
the cut 
\beq
|M(l^\prime\bar{l^\prime} )-\MZ|< 20\GeV.  
\eeq
In order to isolate the $\PW$-boson production, we use instead the transverse 
mass defined as $\MT(l\nu_l)=\sqrt{\ET^2(l\nu_l)-\PT^2(l\nu_l)}$ as the 
physical quantity to be restricted. In the following, we require
\beq\label{addcutsW}
\MT(l\nu_l)< \MW +20\GeV.
\end{equation}
At the tree level, the sensitivity of $\PW\PZ$ production to non-standard 
triple vertices has been studied in detail (see \citere{Haywood:1999qg} and
references therein).
Also the influence of the $\Oas$ QCD corrections on the observability of 
new-physics effects have been extensively analysed 
\cite{Haywood:1999qg,Dixon:1999di,Baur:1995aj}.
The general finding is that the inclusion of anomalous couplings at the 
$\PW\PW\PZ$ vertex 
enhances cross sections and distributions at large values of the partonic 
$\CM$ energy, as well as at large scattering angles of the outgoing bosons. 
Previous calculations \cite{Accomando:2001fn,Kaiser:2004,Meier:2005} have 
shown that
$\Oa$ electroweak corrections to the hadronic di-boson production are sizeable
in exactly this same region. 
In the following, we include the $\EW$ corrections and discuss their effect 
in the analysis of the $\PW\PW\PZ$ triple gauge-boson coupling. We define two 
sample
scenarios, both characterized by large energies and scattering angles in the
di-boson rest frame. The first scenario is fixed by requiring the transverse 
momentum of the reconstructed $\PZ$-boson to be
\beq\label{eq:ptz250}
P_\mr{T}(\PZ) > 250 \GeV. 
\eeq
As a second scenario, we impose the following cut on the transverse 
momentum of any charged lepton
\beq\label{eq:ptl70}
\PT(l)>70\GeV.
\eeq
In these two kinematical regions, we choose to investigate four illustrative 
distributions. We select two energy-like distributions, showing the growth 
with energy of the effects associated to anomalous couplings with respect to 
SM results,
\begin{description}
\item[\qquad$\PTmax(l)$:] maximal transverse momentum of the three
  charged leptons,

\item[\qquad$E(\PZ )$:] energy of the reconstructed $\PZ$-boson,
\end{description}
and two angular distributions
\begin{description}
\item[\qquad$\Delta y(\PZ l)= y(\PZ) - y(l)$:] rapidity difference 
  between the reconstructed $\PZ$-boson and the charged lepton
  coming from the $\PW$-boson decay,
  
\item[\qquad$y(\PZ)$:] rapidity of the reconstructed $\PZ$~boson.
\end{description}
\begin{figure}
  \unitlength 1cm
  \begin{center}
  \begin{picture}(16.,15.)
  \put(-2.5,-1){\epsfig{file=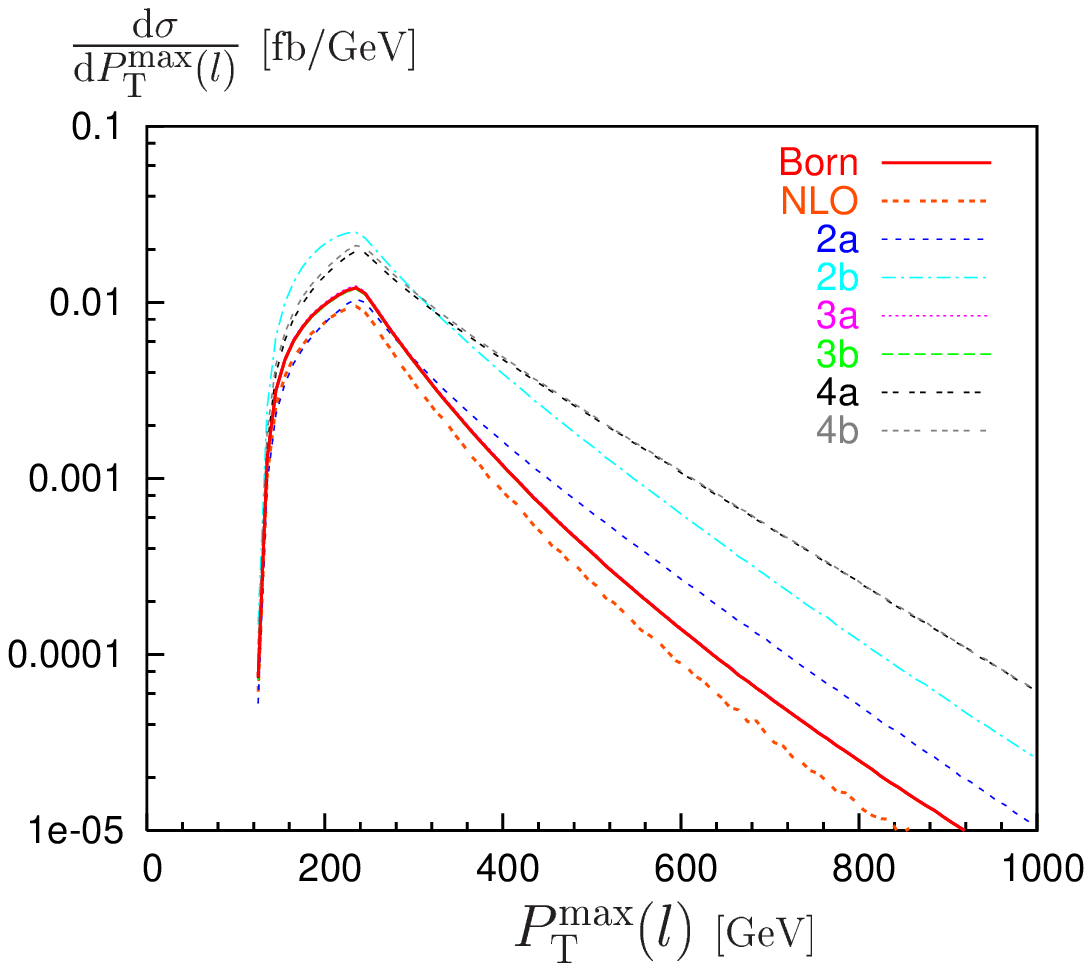,width=14cm}}
  \put(5.5,-1){\epsfig{file=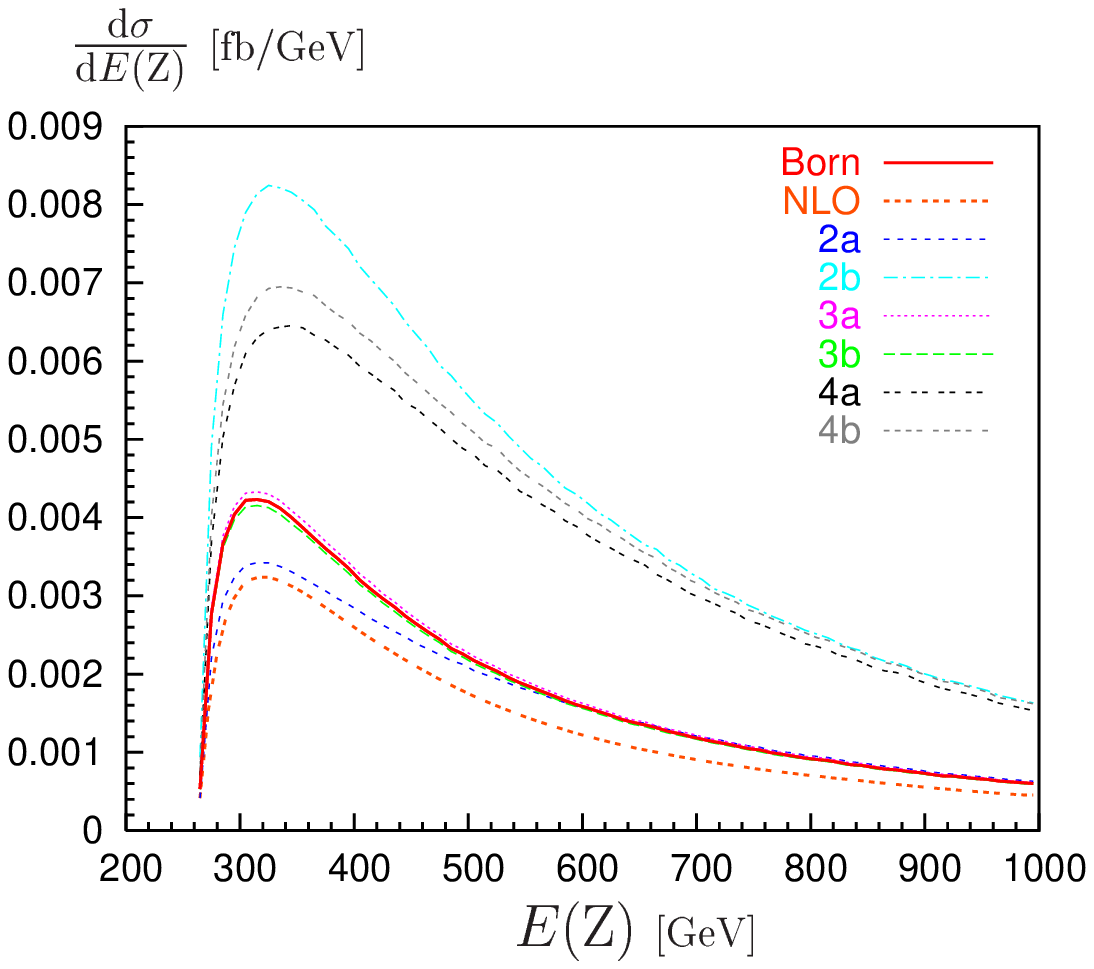,width=14cm}}
  \put(-2.5,-8){\epsfig{file=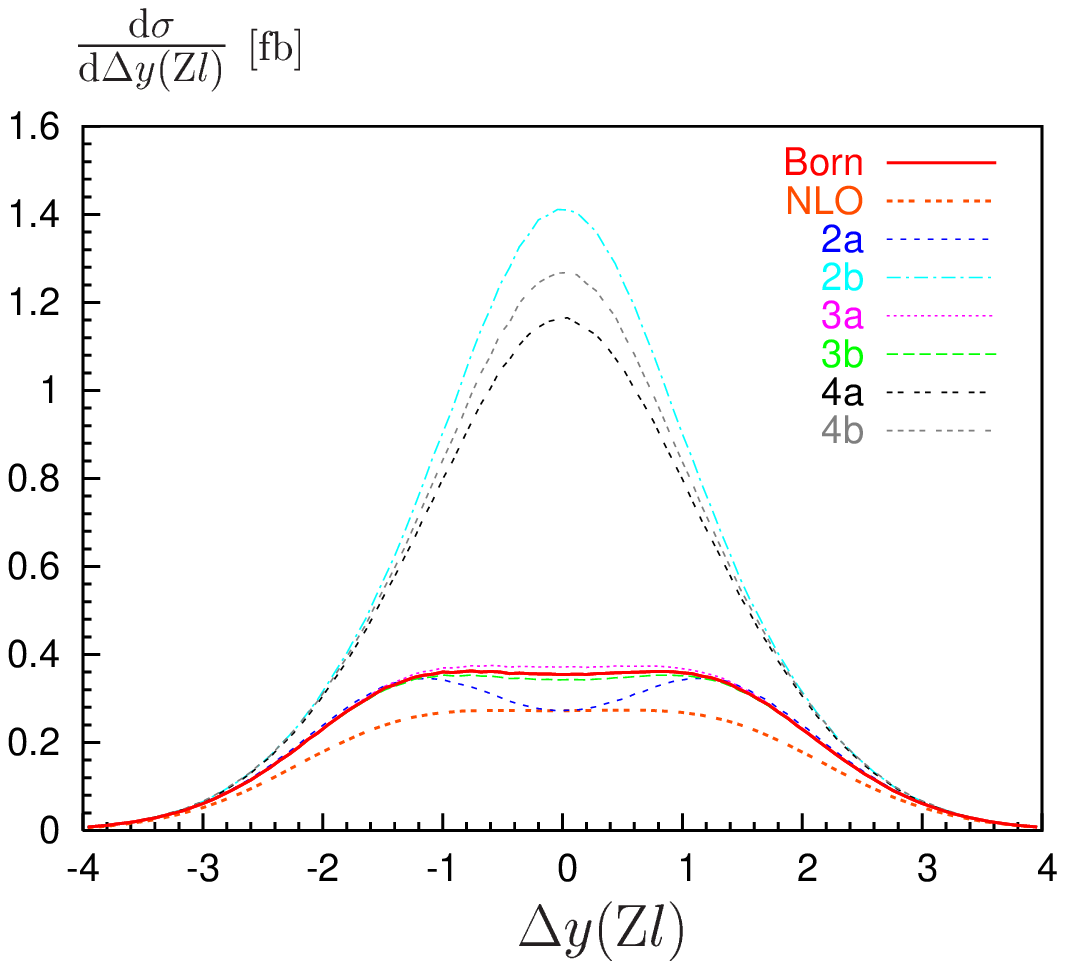,width=14cm}}
  \put(5.5,-8){\epsfig{file=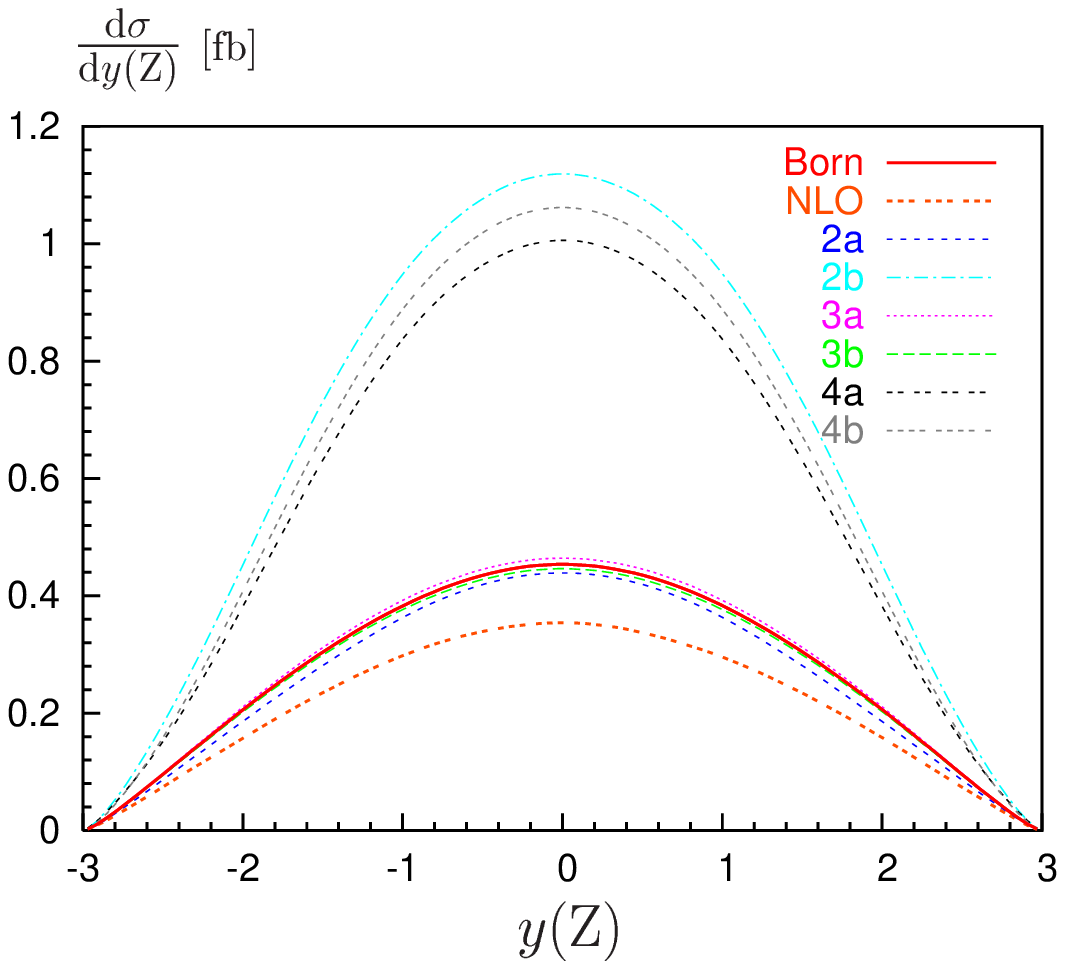,width=14cm}}
  \end{picture}
  \end{center}
\caption{Distributions for $\PW\PZ$ production. 
  (a) Maximal transverse momentum of the charged leptons. (b) Energy of the
  reconstructed $\PZ$-boson. (c) Difference in rapidity between the
  reconstructed $\PZ$-boson and the charged lepton coming from the
  $\PW$-boson decay. (d) Rapidity of the reconstructed $\PZ$-boson.  
  The contributions of the eight final states 
  $l\nu_ll^\prime\bar{l^\prime}$ where $l,l^\prime =e,\mu$
  are summed up, and standard cuts as well as 
  $\PT(\PZ)> 250\GeV$ are applied. Legends as explained in the text. 
}
\label{fi:WZ_s1}
\end{figure}
The rapidity is defined from the energy $E$ and the longitudinal
momentum $P_{\rL}$ by $y=0.5\log ((E+P_\rL)/(E-P_\rL))$.
This latter choice is motivated by a property of the $\PW\PZ$ production.
In the SM, the lowest-order amplitude of the process 
$q_1\bar{q_2}\rightarrow \PW\PZ$ exhibits the well-known approximate 
radiation zero at $\cos\theta^*_\PZ\simeq 0.1 (-0.1)$ for $\PW^+\PZ$ 
($\PW^-\PZ$) production \cite{Baur:1994prl}. Here, $\theta^*_\PZ$ is the 
$\PZ$-boson scattering 
angle with respect to the incoming quark in the di-boson rest frame.
Analogously to the radiation zero in $\PW\gamma$ production, the approximate
amplitude zero in $\PW\PZ$ production can be observed in the distribution
of the rapidity difference $\Delta y(\PZ l)$. At the LHC, the SM at 
leading-order predicts indeed for this observable a dip located at 
$\Delta y(\PZ l)=0$. Radiative corrections and anomalous triple couplings 
might both obscure or enhance this lowest-order SM signature. It is thus 
important to study the interplay between these two contributions.

We start discussing the scenario \refeq{eq:ptz250}. In Fig. 2, we have plotted
the four distributions for the full processes 
$\Pp\Pp\to l\nu_l l^\prime\bar{l^\prime}$ with $l = \Pe$ or $\mu$.
In our notation, $l\nu_l$ indicates both $l^-\bar\nu_l$ and $l^+\nu_l$, \ie 
we sum over the charge-conjugate final states and over all flavours
of the leptons coming from the $\PW$-boson, except $\tau$'s.
The naming of the legend within each plot refers to Table 1.
The upper part of Fig. 2 shows the momentum (left) and energy (right) 
distributions. As one can see, owing to the growth of the non-standard terms 
in the amplitude with the $\CM$ energy, the anomalous couplings give large 
enhancements in the differential cross section at large values of $\PTmax(l)$ 
and $E(\PZ )$. The scenarios $2a/2b$ and $4a/4b$, where $\Delta g_1^\PZ$ and
$\lambda_\PZ$ are different from their SM zero values respectively, give
major deviations from the SM results. This is in agreement with the analysis
of \citere{Baur:1995}. There, it is shown that the associated terms in the 
amplitude grow in fact with the $\CM$ energy squared. In contrast, the terms 
proportional
to $\Delta\kappa_\PZ$ grow only with the $\CM$ energy, thus generating smaller
effects on the cross section. In this specific case, the curves $3a/3b$ in 
Fig. 2 are not distinguishable from the SM result at Born level.  

The $\Oa$ $\EW$ corrections might have an influence on the sensitivity of 
$\PTmax(l)$ and $E(\PZ )$ distributions to triple gauge-boson couplings.
They in fact decrease the lowest-order differential cross section by more than 
20$\%$. Therefore, Born level results overestimate the background rate, 
possibly reducing the sensitivity to new-physics effects. An excess of events 
in the high-energy region could in fact be taken as compatible with the SM 
predictions, and could therefore be obscured or even missed. 

A similar conclusion holds for the two angular distributions shown in the
lower part of Fig. 2. The scenarios $2b$ and $4a/4b$ have the largest impact 
on $\Delta y(\PZ l)$ and $y(\PZ)$ variables. In particular, non-zero 
$\Delta g_1^\PZ$ and $\lambda_\PZ$ values give rise to enhanced positive 
contributions and wash out completely the dip of the approximate radiation 
zero, thus dramatically changing the SM signature. As previously, the $\Oa$ 
$\EW$ corrections affect the afore-mentioned angular observables by a negative 
amount of the order of 20$\%$. The distribution in the rapidity difference 
between the reconstructed $\PZ$-boson and the charged lepton from the 
$\PW$-boson 
decay is also suitable to establish the sign of the non-standard couplings. 
Assuming a positive value for $\Delta g_1^\PZ$ ($2a$ scenario) would generate 
in fact an opposite effect, actually enhancing the SM dip.
Here, the role of the $\EW$ radiative corrections might be subtle. They can 
in fact fake non-standard 
$\Delta g_1^\PZ$ effects, decreasing the lowest-order $\Delta y(\PZ l)$ 
distribution by the same order of magnitude (see the left-side lower plot). 

The role played by the $\EW$ corrections thus depends on 
\begin{figure}
  \unitlength 1cm
  \begin{center}
  \begin{picture}(16.,15.)
  \put(-2.5,-1){\epsfig{file=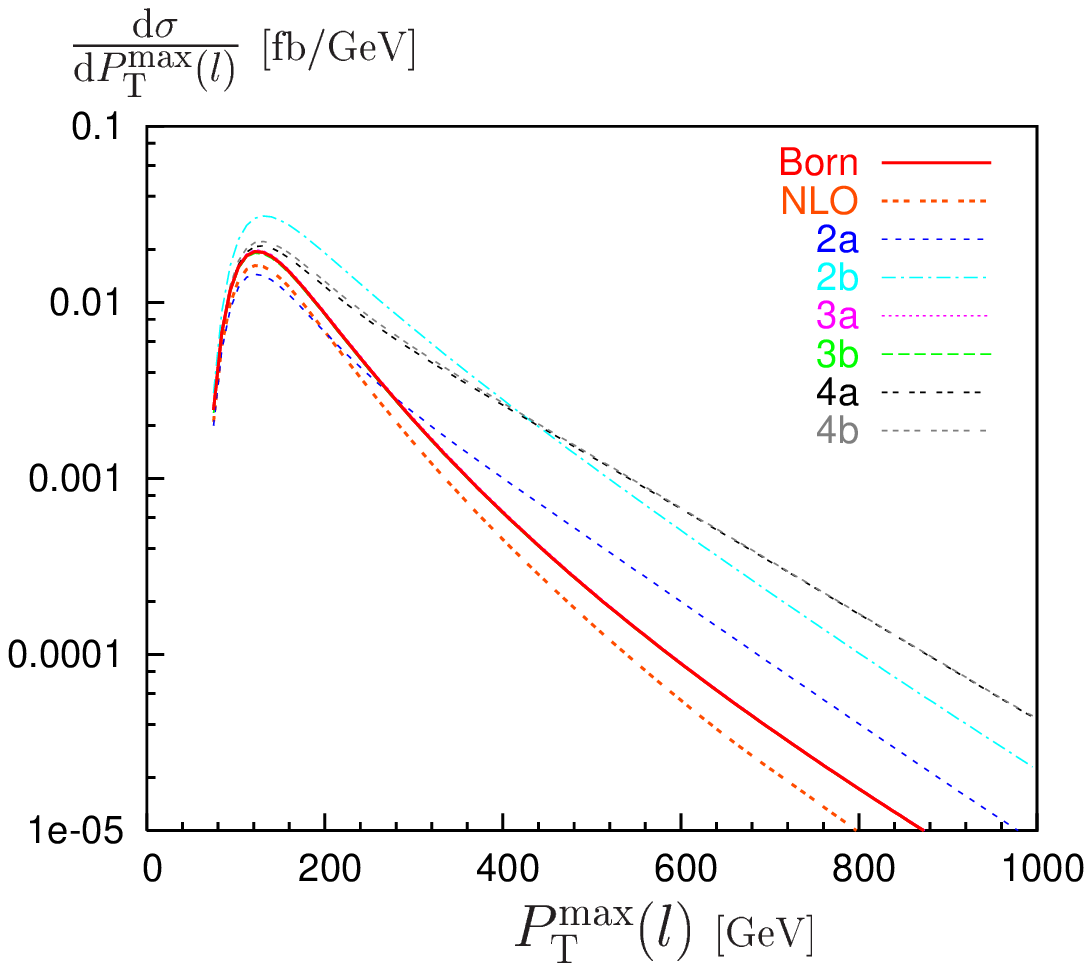,width=14cm}}
  \put(5.5,-1){\epsfig{file=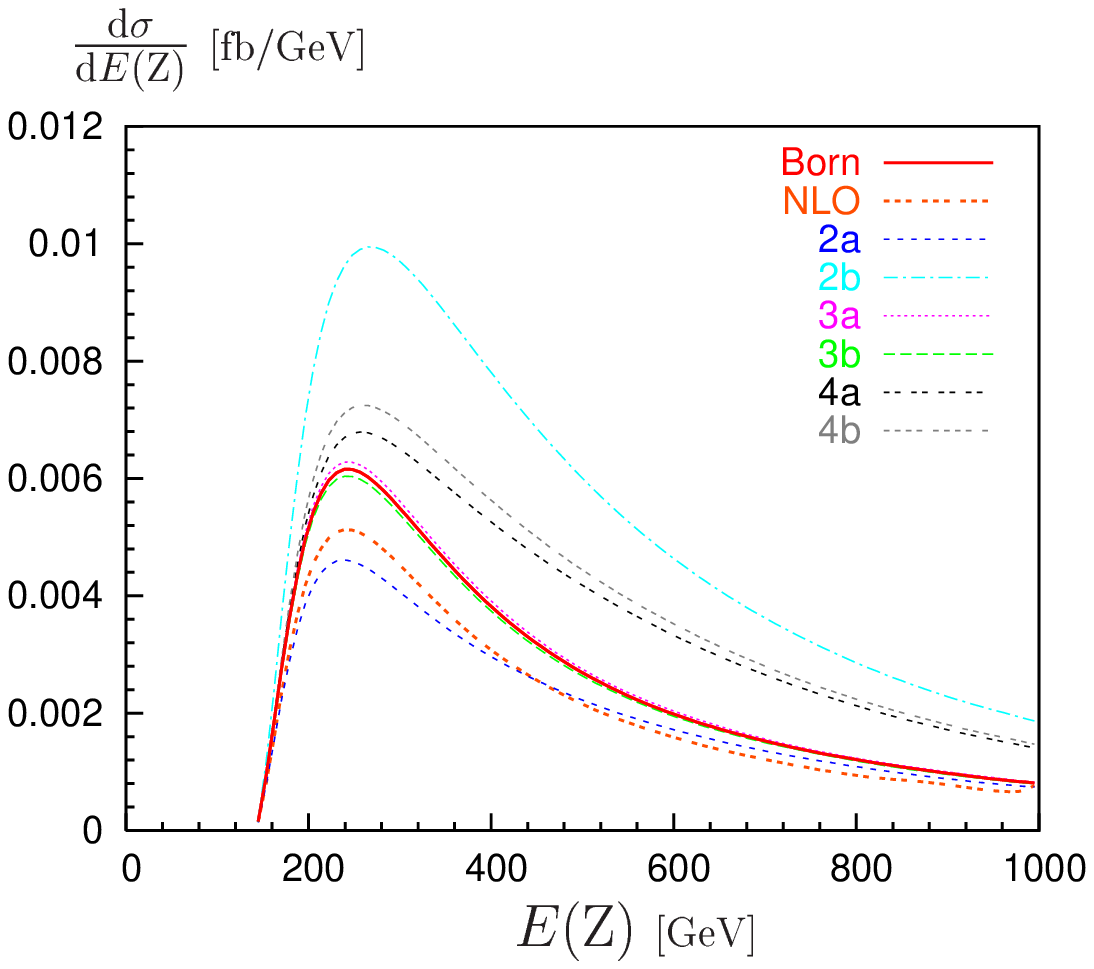,width=14cm}}
  \put(-2.5,-8){\epsfig{file=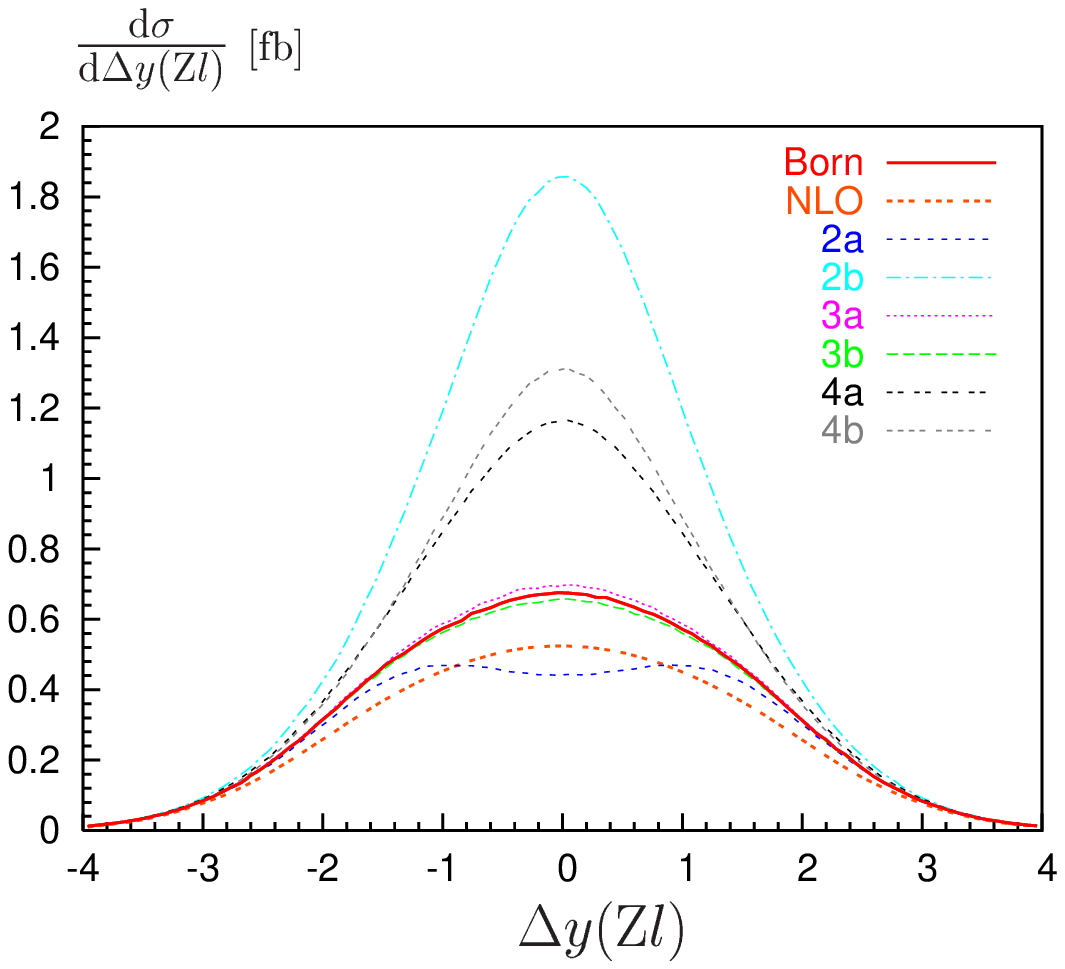,width=14cm}}
  \put(5.5,-8){\epsfig{file=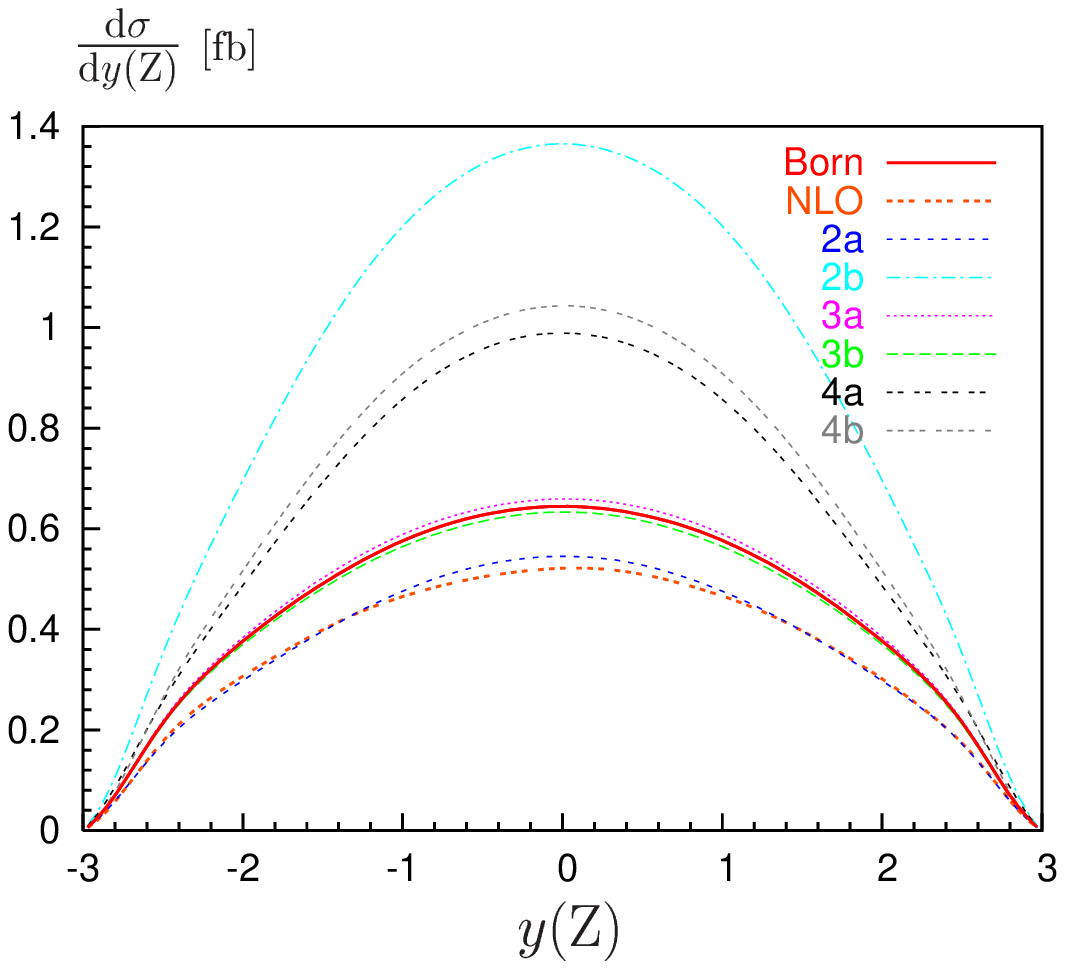,width=14cm}}
  \end{picture}
  \end{center}
\caption{Distributions for $\PW\PZ$ production.
  (a) Maximal transverse momentum of the charged leptons. (b) Energy of the
  reconstructed $\PZ$-boson. (c) Difference in rapidity between the
  reconstructed $\PZ$-boson and the charged lepton coming from the
  $\PW$-boson decay. (d) Rapidity of the reconstructed $\PZ$-boson.  
  The contributions of the eight final states 
  $l\nu_ll^\prime\bar{l^\prime}$ where $l,l^\prime =e,\mu$
  are summed up, and standard cuts as well as 
  $\PT(l)> 70\GeV$ are applied. Legends as explained in the text. 
}
\label{fi:WZ_s2}
\end{figure}
the observable and the scenario at hand. Moreover, it can also vary according 
to the applied kinematical 
\begin{table}
\begin{tabular}{c|c|c|c|c|c|c|c|c|c}
$P_\mr{T}^{\cut}(\PZ)$ & $\Born$ & NLO ($\De [\% ]$) & $2a$ & $2b$ & $3a$ & $3b$ & $4a$ & $4b$ & $[2L\sigma_{\Born}]^{-{1\over 2}}$\\
\hline
250  &  1.672  &  1.296 (-23) &  1.576  &  3.996  &  1.712  &  1.644  &  3.510  &  3.718 & 5.5 $\%$ \\
300  &  0.876  &  0.658 (-25) &  0.940  &  2.496  &  0.896  &  0.862  &  2.366  &  2.478 & 7.6 $\%$\\
350  &  0.490  &  0.354 (-28) &  0.606  &  1.634  &  0.500  &  0.482  &  1.664  &  1.726 & 10.1 $\%$\\
400  &  0.286  &  0.202 (-29) &  0.410  &  1.100  &  0.292  &  0.284  &  1.194  &  1.230 & 13.2 $\%$\\
450  &  0.176  &  0.120 (-32) &  0.286  &  0.756  &  0.178  &  0.174  &  0.866  &  0.888 & 16.9 $\%$\\
500  &  0.110  &  0.074 (-33) &  0.202  &  0.526  &  0.112  &  0.110  &  0.630  &  0.644 &21.3 $\%$\\
\end{tabular}
\caption{Cross sections in $\fba$ for 
$\Pp\Pp\rightarrow l\nu_ll^\prime\bar{l^\prime}$ where $l,l^\prime =e,\mu$
for different cuts (in $\GeV$) on the transverse momentum of the reconstructed 
$\PZ$-boson. All eight final states are summed up, and standard cuts are 
applied.}
\end{table}
cuts. As an example,  
if one considers the kinematical region defined by eq.\refeq{eq:ptl70}, the
similarity between $\Oa$ and non-standard effects is much more evident.
This is shown in Fig. 3 where we plot the same four distributions as before.
Here, NLO SM results and $2a$ scenario display the same behavior as compared 
to the Born SM distributions, independently whether they are energy-like or 
angular-like ($\PTmax(l)$ exhibits this characteristic in the dominant 
low-value range). The deviation from the lowest-order SM results can reach
some tens of per cent in both cases, well exceeding the statistical accuracy. 
The $\EW$ corrections should therefore be taken into account to make sure
that an experimentally observed discrepancy from the Born SM predictions due 
to radiative effects is not misinterpreted as a new-physics signal.    

The advantage of selecting the less stringent kinematical domain
\refeq{eq:ptl70} consists in roughly doubling the statistics, keeping the 
good feature of analysing rather large $\CM$ energies and scattering angles 
to enhance non-standard terms. Taking into account all lepton flavours, one 
has $\sigma_{\Born}({P_\mr{T}(\PZ)>250\GeV})=1.672\fba$ and 
$\sigma_{\Born}({\PT(l)>70\GeV})=2.64\fba$ for scenarios \refeq{eq:ptz250} and 
\refeq{eq:ptl70}, respectively. In these two sample regions, the $\Oa$ 
corrections have similar consequences on the observability of possible 
new-physics effects. In both cases, they are negative and lower the 
lowest-order cross section by about 20$\%$. 

The significance of the $\EW$
corrections can be naively derived from their comparison with the statistical 
error expected at the LHC. In the low luminosity run, they give a 
two-standard-deviation effect (2$\sigma$) with respect to the Born SM 
results. In the high luminosity run, their contribution increases up to 
4-5$\sigma$. The existence of anomalous TGCs might have similar consequences.  
This is illustrated in more detail in Table 2 for the scenario 
\refeq{eq:ptz250}. In columns 3 and 10, we list 
the relative deviation $\De = (\sigma_\NLO-\sigma_\Born )/\sigma_\Born$ and
the statistical accuracy (estimated by taking as a luminosity $L=100
\fba^{-1}$ for two experiments) for some values of the $\PZ$-boson transverse 
momentum cut. We sum over all eight final states $\Pem\Pnebar\mu^-\mu^+$, 
$\Pne\Pep\mu^-\mu^+$, $\mu^-\bar\nu_\mu\Pem\Pep$, $\nu_\mu\mu^+\Pem\Pep$,
$\mu^-\bar\nu_\mu\mu^-\mu^+$, $\nu_\mu\mu^+\mu^-\mu^+$, 
$\Pem\Pnebar\Pem\Pep$, $\Pne\Pep\Pem\Pep$.
This comparison indicates that $\EW$ corrections can be bigger or comparable 
with the experimental precision up to about $P_\mr{T}^{\cut}(\PZ)=500\GeV$. In 
this region the deviation from the Born SM results given by the $\Oa$ 
contributions ranges between $-23$ and $-33\%$. This order of magnitude is 
much larger or at least comparable with the effect of non-standard terms 
coming from $\De g_1^\PZ>0$ 
and $\De\kappa_\gamma$ (see columns 4, 6 and 7 in Table 2). Thus a reliable 
analysis of the afore-mentioned final states requires the inclusion of the 
$\Oa$ $\EW$ corrections. This kind of accuracy is advisable also in a 
low-luminosity run.

\begin{table}
\begin{tabular}{c|c|c|c|c|c|c|c|c|c}
$\Minv^\cut(l\bar{l^\prime})$ & $\Born$ & NLO ($\De [\% ]$) & $2a$ & $2b$ & $3a$ & $3b$ & $4a$ & $4b$ & $[2L\sigma_{\Born}]^{-{1\over 2}}$\\
\hline
500  & 7.239   &  5.559 (-23) & 7.222  &  7.978  & 7.351  & 7.587  & 8.026  &  8.024 & 2.6 $\%$ \\
\end{tabular}
\caption{Cross sections in $\fba$ for 
$\Pp\Pp\rightarrow l\bar\nu_l\nu_{l^\prime}\bar{l^\prime}$ where 
$l,l^\prime =e,\mu$. All four final states are summed up, and standard cuts 
as well as $\Minv(l \bar{l^\prime})> 500 \GeV$ and 
$|\Delta y_{l\bar{l^\prime}}|< 3$ are applied.}
\label{ta:WW_s1}
\end{table}

\subsection{$\PW\PW$ production}

In this section, we discuss the processes 
$\Pp\Pp\to l\bar\nu_l\bar{l^\prime}\nu_{l^\prime}$ ($l,l'=\Pe$ or $\mu$). This
channel contains informations on the charged gauge-boson vertices, 
$\PW\PW\PZ$ and $\PW\PW\gamma$. It can count on the largest cross section 
among all massive vector-boson pair-production processes at the LHC, which 
makes it a favourable channel. Even if it does not allow for a clean and 
unambiguous reconstruction of the two $\PW$ bosons, owing to the presence of 
two neutrinos, it is suitable for measuring triple anomalous couplings. Its 
goodness depends also on the control one can have on the large background from 
$\Pt\bar\Pt$ production. 

We consider the following scenario:
\begin{equation}\label{eq:WWscenarioII}
\Minv(l\bar{l^\prime})> 500\GeV,\qquad 
|\Delta y_{l\bar{l^\prime}}|< 3.
\end{equation}
Possible $\PZ\PZ$ intermediate states are heavily suppressed by the 
invariant-mass cut in \refeq{eq:WWscenarioII}. Therefore, we can safely 
neglect the contributions of $\Pem\Pep\nu_\Ri\bar\nu_\Ri$ 
($\Ri=\mu,\tau$) and $\mu^-\mu^+\nu_\Ri\bar\nu_\Ri$ ($\Ri=\Pe,\tau$) final 
states. We also do not include $\Oa$ corrections to the $\PZ\PZ$ 
intermediate state contributing to the mixed channels 
$\Pp\Pp\rightarrow\Pem\Pep\nu_\Pe\bar\nu_\Pe$ and 
$\Pp\Pp\rightarrow\mu^-\mu^+\nu_\mu\bar\nu_\mu$. 

\begin{figure}
  \unitlength 1cm
  \begin{center}
  \begin{picture}(16.,13.6)
  \put(-2.5,-3){\epsfig{file=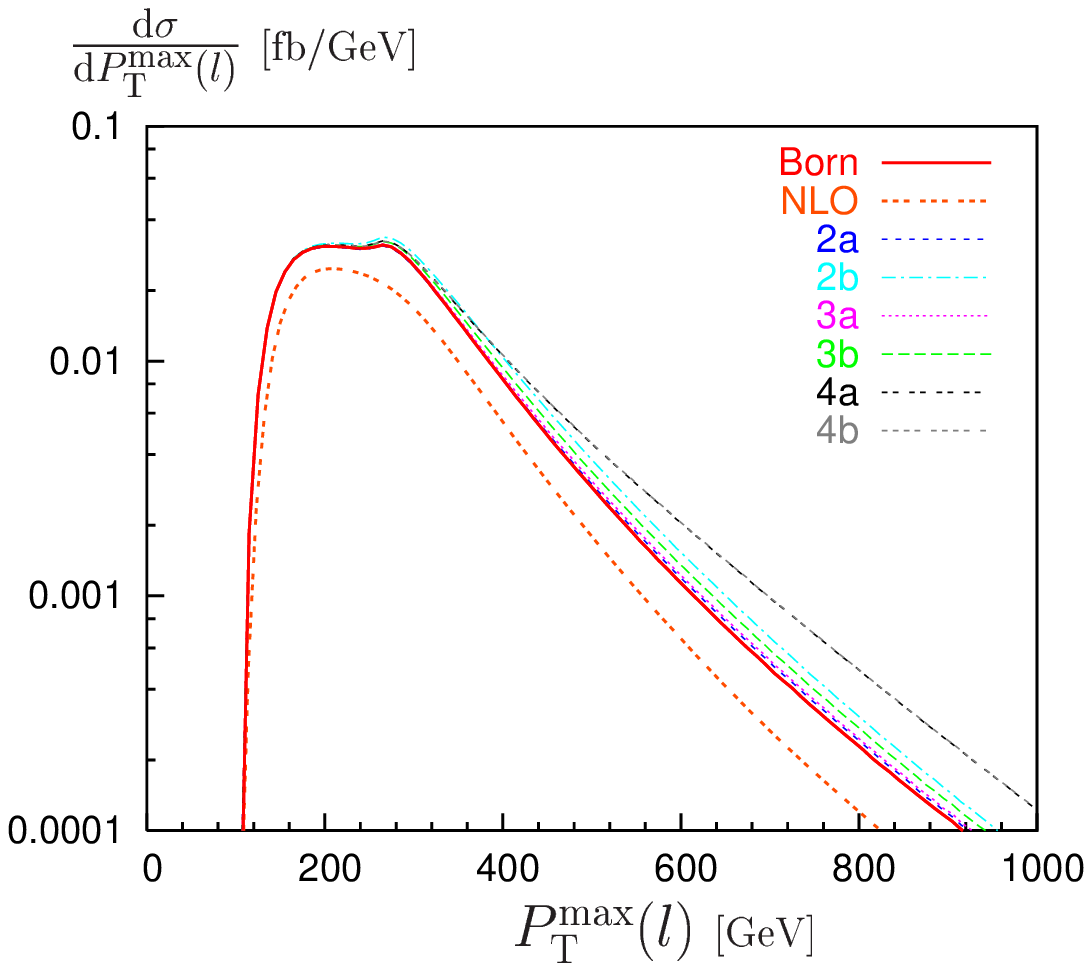,width=14cm}}
  \put(5.5,-3){\epsfig{file=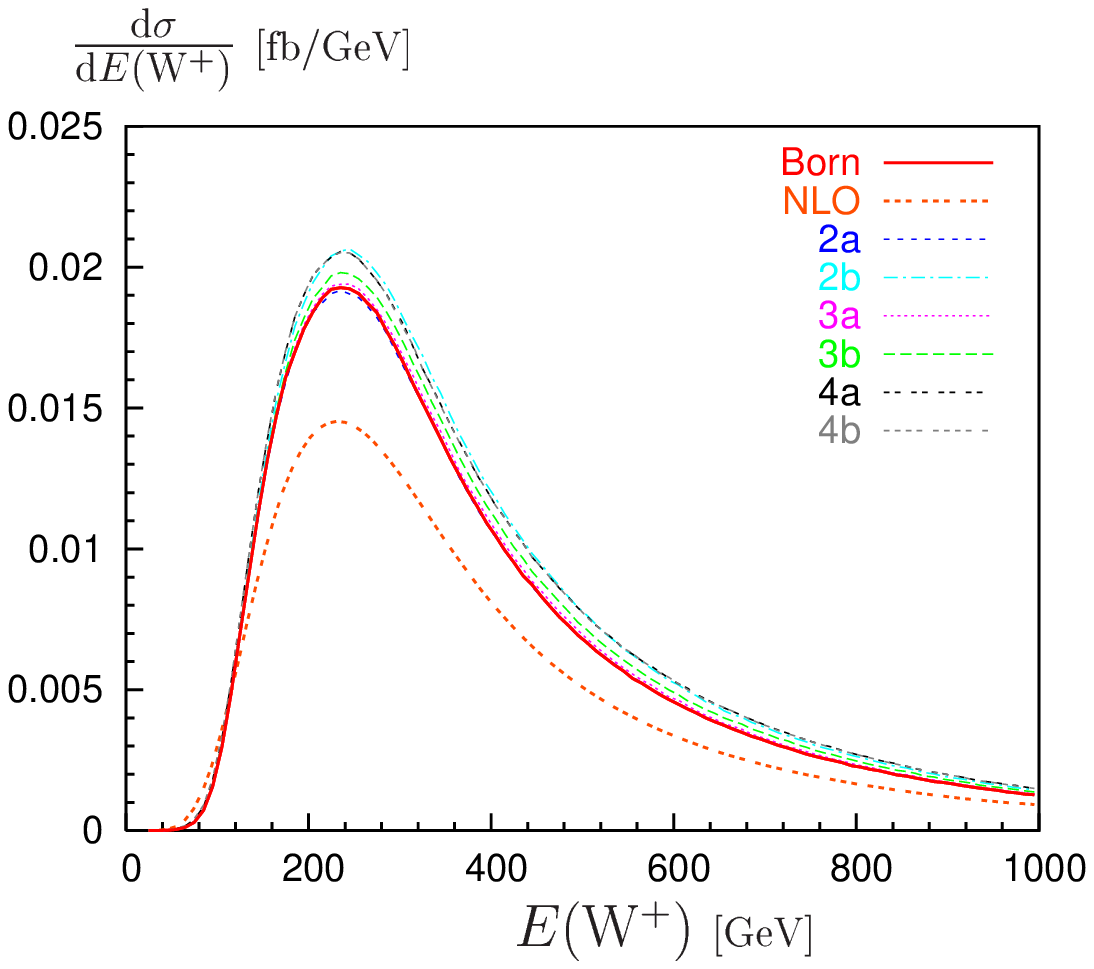,width=14cm}}
  \put(-2.5,-10){\epsfig{file=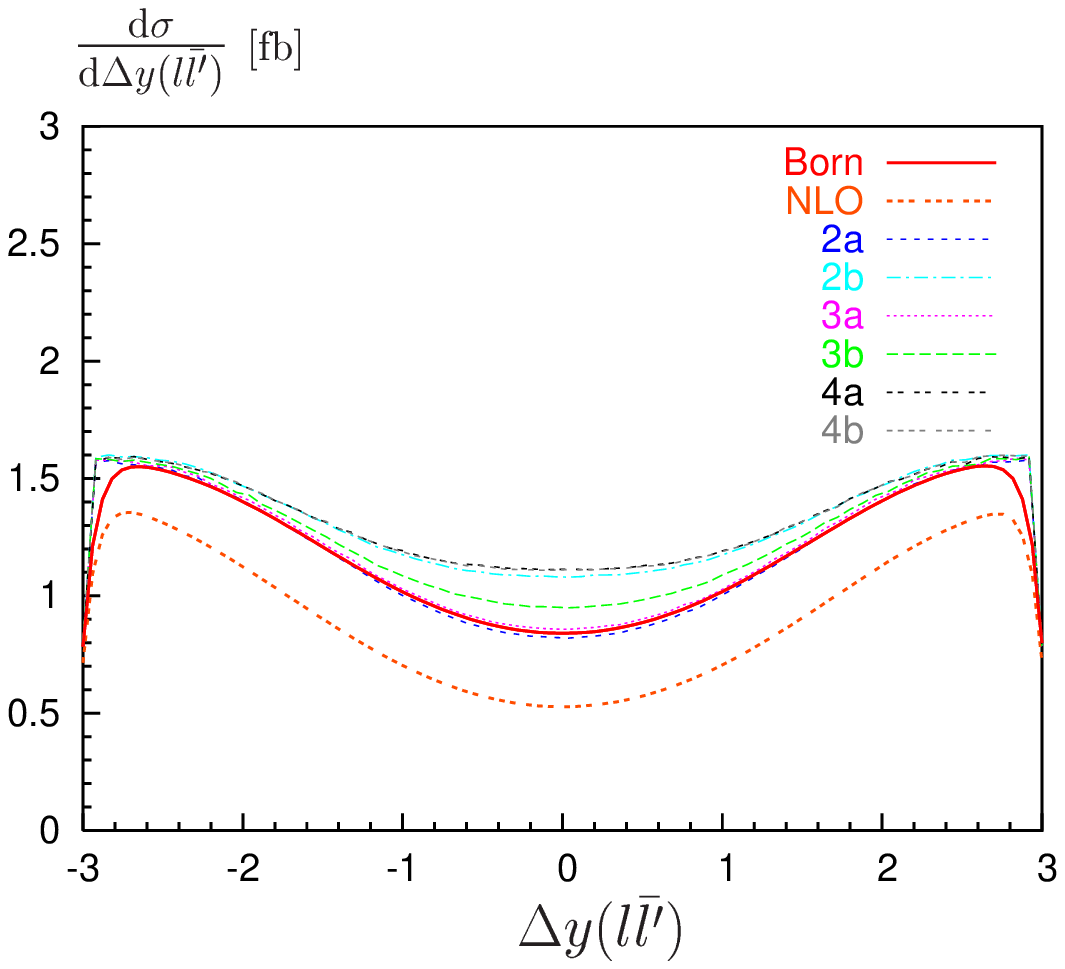,width=14cm}}
  \put(5.5,-10){\epsfig{file=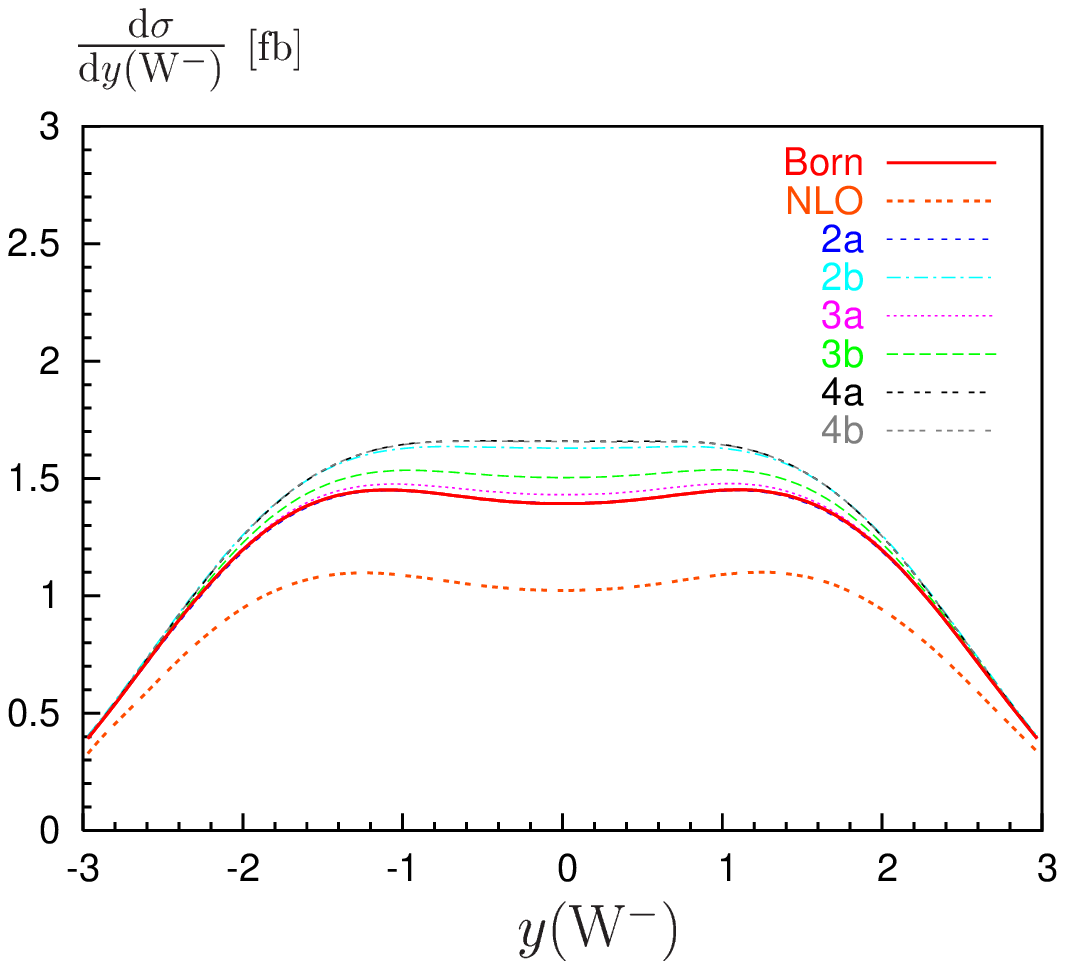,width=14cm}}
  \end{picture}
  \end{center}
\caption{Distributions for $\PW\PW$ production.
  (a) Maximal transverse momentum of the charged leptons. (b) Energy of the 
  $\PW$-boson. (c) Rapidity difference of the two charged leptons. (d) 
  Rapidity of the $\PW^-$ boson. The contributions of the four final states 
  $l\bar\nu_l\nu_{l^\prime}\bar{l^\prime}$ where $l,l^\prime =e,\mu$
  are summed up, and 
  standard cuts as well as $\Minv(l \bar{l^\prime})> 500 \GeV$ and
  $|\Delta y_{l\bar{l^\prime}}|< 3$ are applied. 
  Legends as explained in the text. 
}
\label{fi:WW_s1}
\end{figure}

For $\PW\PW$ production, we choose to discuss distributions in the following 
variables:
\begin{description}
\item[\qquad$\PTmax(l)$:] maximal transverse momentum of the two charged 
leptons,

\item[\qquad$\Delta y(l\bar{l^\prime})$:] rapidity difference between the two
charged leptons,

\item[\qquad$E(\PW^+)$:] energy of the $\PW^+$ boson,

\item[\qquad$y(\PW^-)$:] rapidity of the $\PW^-$ boson.

\end{description}
Despite of the fact that we do not perform a reconstruction of the two $\PW$ 
bosons, the last two unphysical distributions are useful to display some
peculiarities of $\EW$ corrections and anomalous couplings.
In \reffi{fi:WW_s1} we show the four distributions for the final states
$l\bar\nu_l\bar{l^\prime}\nu_{l^\prime}$ ($l,l'=\Pe$ or $\mu$),
with our standard 
cuts applied. The general behaviour of the $\EW$ corrections does not present 
novelties compared to the previous case. As for $\PW\PZ$ production, $\Oa$ 
corrections are in fact enhanced at high $\CM$ energies and large scattering 
angles. This translates into larger radiative effects in the tail of 
transverse momentum and energy distributions, and in the central region of 
rapidity distributions, as shown in the two upper and lower plots of 
\reffi{fi:WW_s1} respectively.

The interest in $\PW\PW$ processes is twofold. The main feature is the 
remarkable statistics of purely leptonic final states.  As shown in 
Table 3, where we sum over the four final states 
$\Pem\bar\Pne\nu_\mu\mu^+$, $\Pne\Pep\mu^-\bar\nu_\mu$,
$\mu^-\bar\nu_\mu\nu_\mu\mu^+$, and $\Pem\bar\Pne\Pne\Pep$, the
estimated experimental precision is around a few per cent at CM
energies above 500$\GeV$. The second characteristic is the stronger interplay
between $\EW$ corrections and anomalous coupling effects. Both total cross 
sections (see Table 3) and distributions exhibit a poor sensitivity
to non-standard terms in $\PW\PW\PZ$ and $\PW\PW\gamma$ vertices. The major 
effects are obtained when the interference between anomalous contributions 
and large SM amplitudes can be exploited. Unfortunately, the $\PW$-boson pair 
production is dominated by the Feynman diagram with t-channel neutrino 
exchange, which does not involve TGCs. The interesting interferences are thus
suppressed \cite{zeppenfeld:1995}. As a result, when looking at the total 
cross section, the effect is at most of order 10\% if compared to the 
lowest-order SM predictions. It slightly increases in some particular 
distributions. 

The optimal case would be considering observables related to the intermediate 
gauge-bosons. As shown in the right-side lower plot of \reffi{fi:WW_s1}, the 
anomalous couplings influence mostly those events where the $\PW$'s are 
produced at large angles with respect to the beam. Unfortunately, for purely 
leptonic final states, gauge-boson variables are not physical as the $\PW$'s 
cannot be reconstructed. One has to resort to observables related to the two
charged leptons in order to find out a measurable effect. This indirect 
detection of the gauge-boson properties might in principle deplete the 
effective strength of the non-standard terms. Selecting appropriate 
variables, like the rapidity difference between the two charged leptons shown 
in the left-side lower plot of \reffi{fi:WW_s1}, their effect can be 
preserved. Here, however, the deviation from the SM result is at most of the 
order of 40\%, and it is concentrated around the dip where the events are 
less abundant. The situation slightly improves if one looks at the 
distribution in the maximum transverse momentum of the two charged leptons.
But still sizeable effects appear only in regions of low statistics.

On the other side, in the same energy domain as defined by 
\refeq{eq:WWscenarioII}, the impact of the $\Oa$ contributions is of much 
grater significance. If one considers the total cross section, it amounts to 
about -23\% of the lowest-order result (see Table 3). For the chosen setup, 
this means a 8$\sigma$ effect which is more than a factor two larger than what
generated by non-standard scenarios.
The distributions plotted in \reffi{fi:WW_s1} confirm this behavior. The $\Oa$ 
effects are in fact shown to be generally bigger than those ones due to 
possible new physics. Thus, for
any decent analysis of the afore-mentioned final states, Monte Carlo programs
should include the electroweak radiative effects. 



\section{Conclusions}
\label{sec:concl}

We have explored some aspects of gauge-boson physics at the LHC, \ie the 
influence of non-standard trilinear gauge-boson couplings on $\PW\PZ$ 
and $\PW\PW$ di-boson production. To this aim, we have analysed two classes of 
processes $\Pp\Pp\rightarrow l\nu_ll^\prime\bar{l^\prime}$ and 
$\Pp\Pp\rightarrow l\bar{\nu_l}\nu_{l^\prime}\bar{l^\prime}$, which contain
$\PW\PZ$ and $\PW\PW$ pairs as intermediate state respectively, and provide
a rather clean leptonic signature. We have examined these processes in the 
physically relevant region of high di-boson invariant mass and large 
vector-boson scattering angle, where effects due to anomalous TGCs are expected
to be maximally enhanced.

In our analysis, we have employed a complete four-fermion calculation, taking 
into account the decays of the gauge bosons as well as the irreducible 
background coming from all not double-resonant Feynman diagrams which give 
rise to the same final state. The primary aim of our study was to understand 
the interplay between the effect due to anomalous TGCs and the influence of 
electroweak radiative corrections. Both contributions to the di-boson 
production processes are enhanced in the kinematical domain of interest.
We have thus compared cross sections and distributions obtained for different 
anomalous TGC parameters with the results predicted by the Standard Model, 
including full $\Oa$ electroweak corrections. The one-loop radiative 
corrections to the complete four-fermion processes have been evaluated in 
double-pole approximation, and keeping leading-logarithmic terms of the ratio 
$\sqrt{\hat{s}}/\MW$ between CM-energy and $\EW$ scale.
In this approximation, the $\Oa$ contribution is split into corrections to the
gauge-boson-pair-production subprocesses, corrections to the gauge-boson 
decays, and non-factorizable corrections.
We have also included the full electromagnetic logarithmic corrections, which
involve the emission of real photons and thus depend on the detector 
resolution.  

In order to illustrate the behaviour and the size of the non-standard TGC 
contributions as compared to the $\Oa$ effects, we have presented various 
cross-sections and distributions. The comparison shows clearly that the $\EW$ 
corrections can be of 
the same order of magnitude and shape as the contributions from the anomalous 
couplings. In the sample scenarios we considered, the $\Oa$ contributions 
decrease the lowest order SM results by $23-33\%$. Their impact thus well 
exceeds the few-per-cent-order statistical error envisaged at the LHC.

As for the majority of the anomalous TGC 
parameters the non-standard terms lead to an increase of the SM results, the 
inclusion of the $\EW$ corrections improves the sensitivity to possible new 
physics by correcting the overestimation of the SM background. In an opposite 
way, when non-standard terms manifest themselves in a decrease of the 
lowest-order results, the $\Oa$ corrections may instead fake anomalous 
contributions. In this case, a pure SM radiative effect could be 
misinterpreted as a new-physics signal.
The $\EW$ radiative effects should therefore be taken into account in 
measuring the $\PW\PW\gamma$ and $\PW\PW\PZ$ vertices at the LHC.   
This conclusion is not peculiar of forseen high luminosities, but applies 
also to the initial low-luminosity run.

\section*{Acknowledgements}
A.~Denner is gratefully acknowledged for extensive discussions on the subject 
treated and for carefully reading the manuscript. We also thank G.~Passarino 
for valuable comments.
This work was supported by the Italian Ministero dell'Istruzione, 
dell'Universit\`a 
e della Ricerca (MIUR) under contract Decreto MIUR 26-01-2001 N.13 
``Incentivazione alla mobilit\`a di studiosi stranieri ed italiani residenti 
all'estero''.


\begin{thebibliography}{9}
\bibitem{Haywood:1999qg}
S.~Haywood, P.~R.~Hobson, W.~Hollik, Z.~Kunszt {\it et al.},
hep-ph/0003275, in {\sl Standard Model Physics (and more) at the LHC},
eds.~G. Altarelli and M. L. Mangano, (CERN-2000-004, Gen\`eve, 2000) p.~117.
\bibitem{Dixon:1999di}
L.~Dixon, Z.~Kunszt and A.~Signer,
Phys.\ Rev.\ D {\bf 60} (1999) 114037
[hep-ph/9907305].

\bibitem{DeFlorian:2000sg}
D.~De Florian and A.~Signer,
Eur.\ Phys.\ J.\ C {\bf 16} (2000) 105
[hep-ph/0002138].

\bibitem{Campbell:1999ah}
J.~M.~Campbell and R.~K.~Ellis,
Phys.\ Rev.\ D {\bf 60} (1999) 113006
[hep-ph/9905386].

\bibitem{Ohnemus:1991gb}
J.~Ohnemus,
Phys.\ Rev.\ D {\bf 44} (1991) 3477.

\bibitem{Frixione:1992pj}
S.~Frixione, P.~Nason and G.~Ridolfi,
Nucl.\ Phys.\ B {\bf 383} (1992) 3.

\bibitem{Baur:1995aj}
U.~Baur, T.~Han and J.~Ohnemus,
Phys.\ Rev.\ D {\bf 51} (1995) 3381
[hep-ph/9410266].

\bibitem{Hollik:2004}
W.~Hollik {\it et al.}, Acta Phys. Polon. B {\bf 35} (2004) 2533 
[hep-ph/0501246].

\bibitem{Accomando:2001fn}
E.~Accomando, A.~Denner and S.~Pozzorini,
Phys.\ Rev.\ D {\bf 65} (2002) 073003
[hep-ph/0110114];
W.~Hollik and C.~Meier, Phys.\ Lett.\ B {\bf 590} (2004) 69 
[hep-ph/0402281].   

\bibitem{Kaiser:2004}
E.~Accomando, A.~Denner, A.~Kaiser, Nucl. Phys. {\bf B706} (2005) 325 
[hep-ph/0409247]; A. Kaiser, dissertation, University Z\"urich 2004.

\bibitem{Meier:2005}
E.~Accomando, A.~Denner, C.~Meier, hep-ph/0509234.

\bibitem{Beenakker:1993tt}
W.~Beenakker {\it et al.},
Nucl.\ Phys.\ B {\bf 410} (1993) 245.

\bibitem{ewee}
M.~Beccaria {\it et al.}, Phys. Rev. {\bf D58} (1998) 093014 
[hep-ph/9805250];\\
P.~Ciafaloni and D.~Comelli,
Phys.\ Lett.\ B {\bf 446} (1999) 278 [hep-ph/9809321];\\
J.~H.~K\"uhn and A.~A.~Penin,
(1999), hep-ph/9906545;\\
M.~Beccaria {\it et al.}, 
Phys.\ Rev.\ D {\bf 61} (2000) 073005 [hep-ph/9906319];\\
V.~S.~Fadin {\it et al.},
Phys.\ Rev.\ D {\bf 61} (2000) 094002 [hep-ph/9910338];\\
W.~Beenakker and A.~Werthenbach, 
Phys.\ Lett.\ B {\bf 489} (2000) 148 [hep-ph/0005316];
Nucl.\ Phys.\ B {\bf 630} (2002) 3 [hep-ph/0112030]; \\
J.~Layssac and F.~M.~Renard,
Phys.\ Rev.\ D {\bf 64} (2001) 053018 [hep-ph/0104205];\\
M.~Melles,
Phys. Rept. 375 (2003) 219 [hep-ph/0104232];\\
J.~H.~K\"uhn {\it et al.},
Nucl.\ Phys.\ B {\bf 616} (2001) 286 [Erratum-ibid. B {\bf 648} (2003) 455]
[hep-ph/0106298];\\
M.~Beccaria, F.~M.~Renard and G.~Verzegnassi, 
Nucl.\ Phys.\ B {\bf 663} (2003) 394 [hep-ph/0304175];\\
S.~Pozzorini, Nucl.\ Phys.\ B {\bf 692} (2004) 135 [hep-ph/0401087];\\
B.~Feucht, J.~H.~K\"uhn, A.~A.~Penin and V.~A.~Smirnov, 
Phys.\ Rev.\ Lett.\ 93 (2004) 101802 [hep-ph/0404082].  



\bibitem{Denner:2001jv} 
A.~Denner and S.~Pozzorini, 
Eur.\ Phys.\ J.\ C {\bf 18} (2001) 461
[hep-ph/0010201].  

\bibitem{Denner:2001gw}
A.~Denner and S.~Pozzorini,
Eur.\ Phys.\ J.\ C {\bf 21} (2001) 63
[hep-ph/0104127].

\bibitem{semitev}
F. Abe {\it et al.} (CDF Collaboration), Phys. Rev. Lett. {\bf 75} (1995) 1017;
F. Abachi {\it et al.} (D0 Collaboration), Phys. Rev. Lett. {\bf 77} (1996) 
3301; Phys. Rev. Lett. {\bf 79} (1997) 1441.

\bibitem{effective_lagrangian}
K.~Hagiwara, R.D.~Peccei, D.~Zeppenfeld, and K.~Hikasa, Nucl. Phys. {\bf B282},
253 (1987).

\bibitem{lep2tgcconvention}
G.~Gounaris {\it et al.}, in {\it Physics at LEP2}, eds. G.~Altarelli, T.~
Sj\"ostrand, F.~Zwirner, CERN 96-01, Vol. 1, pg. 525, hep-ph/9601233.   

\bibitem{pdg}
S.~Eidelman {\it et al.}, Phys. Lett. {\bf B592}, 1 (2004); and 2005 partial 
update for the 2006 edition available on the PDG WWW pages 
(URL: http://pdg.lbl.gov/).

\bibitem{formfactor}
U.~Baur and D.~Zeppenfeld, Phys. Lett. {\bf B201}, 383 (1988); Nucl. Phys. 
{\bf B308}, 127 (1988).

\bibitem{Denner:2000bj}
A.~Denner, S.~Dittmaier, M.~Roth and D.~Wackeroth,
Nucl.\ Phys.\ B {\bf 587} (2000) 67
[hep-ph/0006307].

\bibitem{Berends:1981uq}
F.~A.~Berends, R.~Kleiss, P.~De Causmaecker, R.~Gastmans, W.~Troost and T.~T.~Wu,
Nucl.\ Phys.\ B {\bf 206} (1982) 61;
%
R.~Kleiss,
Z.\ Phys.\ C {\bf 33} (1987) 433.

\bibitem{Hagiwara:pw}
K.~Hagiwara {\it et al.}  [Particle Data Group Collaboration],
Phys.\ Rev.\ D {\bf 66} (2002) 010001.

\bibitem{mtop}
The CDF Collaboration, the D0 Collaboration, the Tevatron Electroweak Working 
Group, hep-ex/0404010.

\bibitem{Hocker:2001xe}
A.~H\"ocker, H.~Lacker, S.~Laplace and F.~Le Diberder,
Eur.\ Phys.\ J.\ C {\bf 21} (2001) 225
[hep-ph/0104062].

\bibitem{cteq}
J.~Pumplin, D.R.~Stump, J.~Huston, H.L.~Lai, P.~Nadolsky and W.K.~Tung, 
JHEP 0207 (2002) 012 [hep-ph/0201195].

\bibitem{Baur:1994prl}
U.~Baur, T.~Han and J.~Ohnemus,
Phys.\ Rev.\ Lett. {\bf 72} (1994) 3941 
[hep-ph/9403248].

\bibitem{Baur:1995}
H.~Aihara {\it et al.}, hep-ph/9503425.

\bibitem{zeppenfeld:1995}
D.~Zeppenfeld, hep-ph/9506239.

\end{thebibliography}
\end{document}